\documentclass[twocolumn,a4paper,usenatbib]{mnras}
\usepackage{graphicx}
\usepackage{psfrag}
\usepackage{amsmath}
\usepackage{amssymb}
\usepackage{color,xcolor}
\usepackage{hyperref}
\hypersetup{
    colorlinks=true,
    citecolor=blue,
    filecolor=black,
    linkcolor=blue,
    urlcolor=magenta,
    linktocpage=true,
    breaklinks = true
}

\usepackage{subfigure,float}
\usepackage{multirow}
\usepackage{makecell}
\usepackage[normalem]{ulem}

\usepackage{array,booktabs}
\usepackage{siunitx}

\usepackage[authoryear]{natbib}

\def\aj{Astron. J.}
\def\apj{Astrophys. J.}
\def\apjl{Astrophys. J.}
\def\mnras{Mon. Not. Roy. Astron. Soc.}

\def\aap{Astron. Astrophys.}
\def\apjs{Astrophys. J. Suppl.}
\def\nat{Nature}
\def\physrep{Physics Reports}
\def \pasj {Publ. Astron. Soc. Jap.}

\newcommand{\beq}{\begin{equation}}
\newcommand{\eeq}{\end{equation}}

\newcommand{\ber}{\begin{eqnarray}}
\newcommand{\eer}{\end{eqnarray}}

\def \ff {${\rm FF}$}

\def \mpc {\rm Mpc}
\def \mpch {\rm h^{-1}Mpc}

\def \dth {D_{\rm th}} 
\def \dthc {D^C_{\rm th}} 
\def \dthmin {D_{\rm th} ^{\rm min}}
\def \dthmax {D_{\rm th} ^{\rm max}}
\def \nc {$N_{\rm Scl}$} 
\def \ng {$N_{\rm gal}$} 
 
\def \nscl {$N_{\rm Scl}$}
\def \ncell {$N_{\rm cell}$}
\def \nr {$N_{\rm R}$}
\def \scl {supercluster~}
\def \scls {superclusters~}

\def \Scls {Superclusters~}

\def \vth {V_{\rm th}}

\title[Morphology of the superclusters]{The shape distribution of superclusters in SDSS DR12}

\author[Bag, Liivamägi \& Einasto]{Satadru Bag,$^1$\thanks{satadru@kasi.re.kr}  Lauri Juhan Liivamägi,$^{2}$\thanks{juhan.liivamagi@ut.ee} Maret Einasto$^2$
\\
\\
$^{1}$ Korea Astronomy and Space Science Institute, Daejeon 34055, Korea\\
$^{2}$ Tartu Observatory, University of Tartu, Observatooriumi 1, 61602 Tõravere, Estonia
}

\date{}

\pubyear{2021}

\begin{document}
\label{firstpage}
\pagerange{\pageref{firstpage}--\pageref{lastpage}}
\maketitle

\begin{abstract}
Galaxy superclusters, the largest galaxy structures in the cosmic web, are formed due to the gravitational collapse (although they are not usually gravitationally bound). Their geometrical properties can shed light on the structure formation process on cosmological scales, hence on the fundamental properties of gravity itself. In this work we study the distributions of the shape, topology and morphology of the \scls extracted from Sloan Digital Sky Survey Data Release 12 (SDSS DR12) main galaxy sample and defined in two different ways -- using fixed and adaptive density threshold in the luminosity-density field. To assess the geometry and topology of each individual supercluster, we employ Minkowski functionals and Shapefinders, precisely calculated by the shape diagnostic tool SURFGEN2. Both supercluster samples produce similar shape distributions. Perhaps not surprisingly, most superclusters are spherical in shape with trivial topology. However, large superclusters with volumes $ V \gtrsim 10^{4}$~Mpc$^{3}$ are statistically found to be filamentary with non-zero genus values. The results, supercluster catalogues and shape distributions are publicly available.
\end{abstract}


\section{Introduction}
The large-scale distribution of matter in the Universe is the result of collapse 
of small initial Gaussian fluctuations in density 
\citep[see e.g.][]{1985ApJ...292..371D,1995PhR...262....1S,
Bernardeau:2001qr,2005Natur.435..629S,Springel:2006vs}. 
Nonlinear gravitational amplification converts these isotropic random fluctuations 
into a field with striking geometrical structure: 
empty 3-dimensional voids bounded by 2-dimensional sheets, 
which intersect in filaments, with the filaments joining at nodes 
forming clusters 
\citep[e.g.][]{1978MNRAS.185..357J,1978ApJ...222..784G,1982Natur.300..407Z,
1984MNRAS.206..529E,1986ApJ...302L...1D,MELOTT19901,1996Natur.380..603B,
1996ApJ...462L...5S,2010MNRAS.409..156B,Jasche:2014vpa}.  The study of this ‘cosmic web’ is important in modern cosmology. 
The history of galaxy formation in different regions 
of the web have been affected in different ways, giving different 
observational tests for models for the assembly of galaxies. 
On a more general scale, the cosmic web can be used to test 
the models for gravity and cosmology pertaining to both linear 
and non-linear structure formation.

The largest structures in the cosmic web are galaxy
superclusters, overdensity regions which consist of galaxy groups and clusters, 
connected by filaments. Superclusters play a key role in understanding the properties and evolution of the cosmic web.
Examples of rich superclusters that have been studied in detail in literature include, for example, the Shapley supercluster
\citep{1930BHarO.874....9S}, the Corona Borealis supercluster
\citep[][and references therein]{2021A&A...649A..51E}, Laniakea supercluster \citep{tully:2014},
the supercluster SCl~A2142 \citep{2015A&A...580A..69E}, and the Saraswati supercluster \citep{2017ApJ...844...25B}, etc.
In some cases, rich superclusters are close together and form supercluster
complexes, as the Sloan Great Wall \citep{2004ogci.conf....5V, 2016A&A...595A..70E}
and the BOSS Great Wall \citep{2016A&A...588L...4L, 2022arXiv220408918E}.
Superclusters can be defined in a different ways, recently briefly
reviewed by \citet{Einasto2020}.  

In the present study we adopt the traditional definition of
superclusters as high-density regions in the cosmic web. We use galaxy data from the 
Sloan Digital Sky Survey (SDSS) to calculate the luminosity-density field
and define connected high-density regions in this field as galaxy superclusters.
With this method the choice of the density level to determine superclusters is
important. We use two approaches for that. One way is to use fixed threshold luminosity-density
limit. At low threshold density superclusters merge to form very large connected systems,
and at very high threshold density only the highest density cores belong to superclusters.
\citet{Einasto:2011zc} proposed that, using data on superclusters from the richest galaxy system in the
nearby Universe, the Sloan Great Wall, the threshold density to define individual
superclusters should be close to $\dth=5.0$ in the units of mean density. We apply this threshold density also in the 
present paper. Another way to choose the density level for is to find the threshold value for each supercluster individually, based on the density distribution around the location of a supercluster.
We describe this method in detail below.

The large-scale structure datasets are analysed using numerous statistics, as recently reviewed in \citet{Libeskind18}, however the focus has been on the  correlation statistics. 
With the major advances in the cutting-edge optical and radio surveys in recent years, well complemented by the recent progresses in large-scale simulations, it has now become extremely important to look beyond the simplest 2-point correlation statistics (or power spectrum) and search for methods which can extract more information out of the persistently growing datasets. 
However, the higher order functions can be difficult to calculate, and sometimes they face conceptual challenges. 
In this context, the Minkowski functionals (MFs) provide a robust and powerful means of studying the morphology of the large-scale
structure of the Universe.
They implicitly contain information from all of the N-point correlation functions and play a role that is complementary to that of the correlation statistics in probing the gravity at non-linear cosmological scales. 
In view of the upcoming spectroscopic and photometric surveys like DESI, LSST, the importance of MFs in analysing large-scale structure datasets is expected to grow rapidly in the near future. 
Since MFs were introduced in cosmology by \cite{mecke}, they have been frequently employed to study the morphology of the cosmic web \citep{Schmalzing1997, Sahni:1998cr, Sathyaprakash:1998gy, Hikage2003,M2007A, Einasto:2011,Pratten, Wiegand:2013xfa, Wiegand_2017, Matsubara:2020fet, Matsubara:2020knr, Lippich:2020vpy,Appleby:2021xoz}, the cosmic reionization process \citep{Friedrich2011,yoshiura:2017, Bag:2018zon, Chen:2018enj, Kapahtia:2021eok, Pathak:2022ahj}, as well as the cosmic-microwave background (CMB) \citep{Schmalzing1998,Novikov1999,Novikov2000,Hikage2006}. 
With few exceptions, these studies calculate `global' 
MFs of the whole isodensity surface.

In this paper 
our approach has two important differences from the majority of the previous studies of MFs in the literature.
First, instead of studying the global MFs of the whole volume, 
we explore the morphology in terms of `shape' of the individual superclusters. 
Previously, a number of studies explored the morphology and evolution of a small number of
superclusters  \citep{M2007A, M2011c, M2014, J2019, J2021}. Our present study extends
morphology analysis of superclusters, for the first
time, to a large dataset. 
Secondly,
instead of focusing on just MFs, in this work we study their ratios 
which are introduced as `Shapefinders' by \citet{Sahni:1998cr}. 
The Shapefinders provide us with the measure of the three physical 
dimensions of an object (thickness, width, and length of a cluster or void or, in our case, of a supercluster). 
Our main method of analysis shall be the computationally advanced version of the SURFGEN algorithm \citep{Sheth:2002rf, Sheth:2006qz}. 
This algorithm
provides means of determining 
the geometrical and topological properties of isodensity contours delineating 
superclusters and voids within the cosmic web \citep{Sheth:2003vm,Shandarin:2003tx}. 
We apply SURFGEN2 algorithm which models the surfaces of the clusters and voids through the advanced `Marching Cube 33' triangulation scheme \citep{marcube,mar33}. 
Its accuracy in determining MFs and Shapefinders is much better than the traditional methods \citep{Schmalzing1997}, such as the techniques based on Crofton’s formula \citep{Crofton_1868}, 
Koenderink’s invariant \citep{Koenderink} or germ-grain models \citep{mecke, Schmalzing:1995qn, Wiegand:2013xfa}.

The paper is organised as follows. Section~2 describes how we construct the density field using the observed galaxies in SDSS DR12. Next, in Section~3 we study the density field using percolation and define the superclusters using isodensity surfaces. A brief overview of MFs and Shapefinders (as well as the code SURFGEN2) is provided in Section~4. We present our main result in Section~5 where we study the shape of \scls defined using two approaches: using a globally fixed or flat density threshold and with individual thresholds set using an adaptive mechanism. We summarise our results and conclude in Section~6. Throughout this paper, we assume the cosmological parameters provided in \citet{Planck:2016VIII}: the Hubble constant $H_0 = 67.8\, {\rm km\, s}^{-1}\, \mpc^{-1}$, the matter density $\Omega_\mathrm{m} = 0.308$, and the dark energy density $\Omega_\mathrm{\Lambda} = 0.692$.

\section{SDSS DR12 main galaxy sample and constructing the luminosity density field}

\subsection{Galaxy data}
Galaxy superclusters in this study are based on the main contiguous area sample from SDSS DR12, the so-called Legacy Survey \citep{Eisenstein:2011sdss,Alam:2015dr12}.
Galaxy data in the more recent SDSS data release 16 is virtually unchanged compared to the DR12 in the Legacy Survey area \citep{Ahumada:2020dr16}.
Specifically, we use the value added catalogue of galaxies, groups and clusters by \citet{Tempel:2017gr}. 
The galaxy sample is volume-limited and consists of $584\,449$ galaxies with spectroscopic redshifts up to $z=0.2$ and brighter than the Petrosian $r$-band magnitude 17.77. 
The catalogue also contains $88\,662$ galaxy groups with a minimum of two members, assembled using a modified friend-of-friend (FoF) method with a variable linking length. 
Details about compiling the galaxy sample are given in \citet{Tempel:2014gr} and \citet{Tempel:2017gr}.

\subsection{Luminosity density fields}
The density field from the SDSS DR12 main galaxy sample was constructed closely following \cite{Juhan_thesis, Liivamagi:2010jg}. 
In short, galaxy luminosities are interpolated onto a regular Cartesian grid with a three-dimensional smoothing kernel. 

We use two properties given in the \citet{Tempel:2017gr} galaxy catalogue to reduce observational effects on the density field.
First, we use the `corrected' distance, which takes into account the velocity dispersion in groups and clusters suppressing the cluster-finger (finger-of-god) redshift distortions \citep{Kaiser:1987qv}.
This correction, of course, is only statistical and not applicable to determine the positions of individual galaxies, however, one can make an assumption that clusters and groups should be more or less spherical and not strongly elongated along the line of sight.

Another important issue is that the number of galaxies in volume-limited samples decreases with distance due to only brighter objects being observed.
In order to counter that, galaxy luminosities are weighted before interpolation, this compensates for the amount of unobserved luminosity, i.e. galaxies that fall outside the observational magnitude limits.
The weight is calculated as the ratio between the area under full galaxy luminosity function and the fraction that is inside the survey visibility window at the distance of the galaxy.
As with the cluster-finger suppression, doing this is correct only in a statistical sense.
In this work, we are looking at structures on much larger scales than individual galaxies and therefore applying both of these `corrections' is justified.
With these procedures, the resulting density field is cleaned from the smearing effect of the cluster-fingers, and at the same time, retains roughly constant value with the distance \citep{Liivamagi:2010jg}, despite the falling number density of galaxies.

\begin{figure}
\centering
\includegraphics[width=\linewidth]{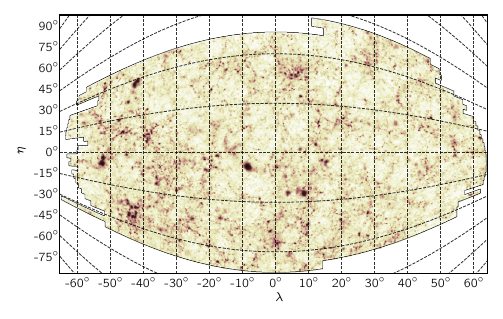}
\caption{Sky projection of the density field of the SDSS DR12 main galaxy sample in survey coordinates $\lambda$ and $\eta$. Black lines show the outline of the mask.}
\label{fig:SDSSDF}
\end{figure}

Luminosity densities $\ell$ on the grid are calculated as
\begin{equation}
    \ell ({\bf r}_{\bf i}) = \sum_{\rm gal} K_B^{(3)} \left({\bf r}_{\rm gal} - {\bf r}_{\bf i}; a, \Delta\right) L_{\rm gal, w},
\end{equation}
where ${\bf r}_{\rm gal}$ are the coordinates and $L_{\rm gal, w}$ is the weighted luminosity of a galaxy. 
The three-dimensional smoothing kernel $K_B^{(3)}$ is a direct product of $B_3$ spline functions (see Appendix~\ref{app:kern} for details)
\begin{equation}
    \begin{array}{l}
    K_B^{(3)}({\bf r};a,\Delta) = \left(\frac{\Delta}{a}\right)^3 B_3(x/a)B_3(y/a)B_3(z/a),\qquad~ \\ \hfill {\bf r} \equiv (x,y,z).
    \end{array}
\end{equation}
Parameters $a$ and $\Delta$ denote smoothing scale and lattice cell length respectively.
Galaxy data and resulting density field are given in a rectangular coordinate system based on the SDSS survey coordinates $\eta$ and $\lambda$, and comoving distance $d$:
\begin{equation}
\begin{array}{l}
    x = -d\, \sin\lambda,\\[-1pt]
    y = d\, \cos\lambda \cos \eta,\\[-1pt]
    z = d\, \cos\lambda \sin \eta.
\end{array}
\label{eq:xyz}
\end{equation}
We have chosen the grid cell length to be $1~\mpc$, sufficiently smaller than the structures we are studying, resulting in a lattice with $N = 1069 \times 890 \times 1542=1\,467\,074\,220$ grid points. 
The kernel used for smoothing has an effective radius of $a=12 ~\mpc$, which roughly corresponds to $8 ~\mpch$ used in \citet{Liivamagi:2010jg}. 
Smoothing scale is essentially a free parameter, however, our chosen value roughly describes the transition scale from linear to non-linear regime in structure formation \citep[e.g.][]{1990eaun.book.....K,2011iteu.book.....G}.
As the grid is rectangular while the galaxy sample has a cone shape, there are many lattice points in the volume with no data.
Therefore we apply a mask that closely follows the sample footprint on the plane of the sky (see Figure~\ref{fig:SDSSDF}), as well as having front and back faces on set distances.
For convenience we convert the luminosity densities into the units of average density $D(\bf r)=\ell(\bf r)/\bar{\ell}$.
Mean density $\bar{\ell}$ is found as the sum of all densities in the `unmasked region' divided by its volume.

The number of grid points in the unmasked region is $N_{\rm unmasked}=418\,879\,964$.
This number is also the unmasked/observed volume ($V_{\rm obs}$) in the units of $\mpc^3$, which is the effective volume of one grid cell. 
Only the unmasked cells of the density field are used in analysis. 

\section{Number of superclusters and percolation}

We identify the isolated regions above a given density threshold $\dth$ using the friend-of-friend algorithm on the grid with the linking length of one grid cell length, i.e. linking adjacent vertices; we refer to these as `isolated overdense regions'. 
In this section we study the number of such regions (\nr) and their percolation properties using volume fractions -- `filling factor' (FF) and  `largest cluster statistics' (LCS)\footnote{While calculating FF and LCS in this work, we simply count the grid cells inside each region instead of calculating the volume using SURFGEN2 (because counting cells in a large region would be a good measure of the actual volume while doing the cluster statistics, and the latter method is computationally expensive)} -- which are defined as follows \citep{Yess1996,Shandarin:1997fc}: 
\begin{align}
    {\rm FF} &= \frac{\rm total ~volume ~of ~all ~the ~isolated ~overdense ~regions}{{\rm volume ~of ~the ~whole ~unmasked ~region}~ (=V_{\rm obs}) }\;, \\
    {\rm LCS}&=\frac{\rm volume ~of ~the ~largest ~isolated ~overdense ~region}{\rm total ~volume ~of ~all ~the ~isolated ~overdense ~regions} \;.
\end{align}
To put it simply, FF and LCS measure the fraction of sample volume that is overdense, and the fraction of overdense volume covered by the largest region respectively.
\begin{figure}
\centering
\includegraphics[width=\linewidth]{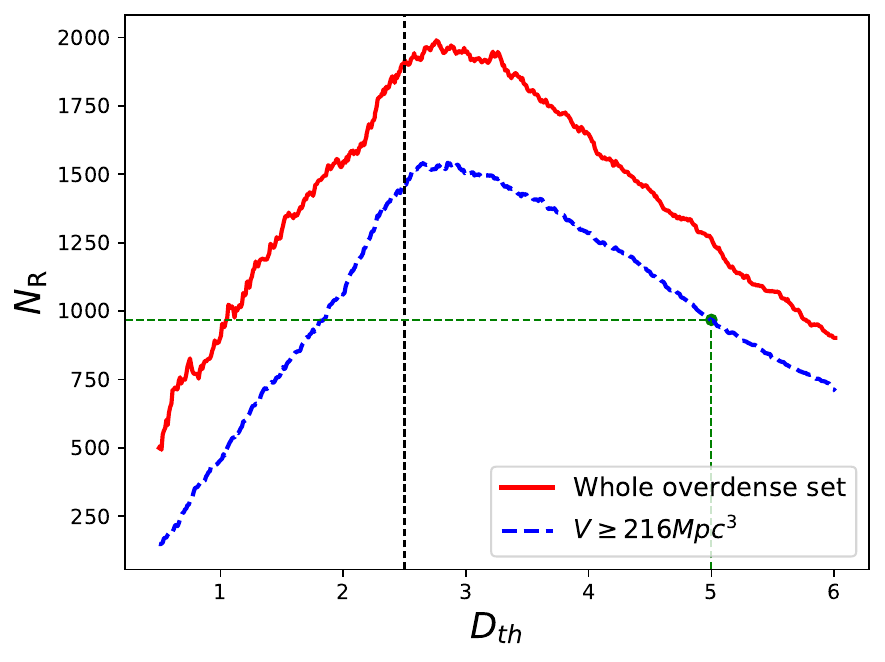}
\caption{Number of isolated overdense regions (\nr) as a function of the density threshold $\dth$. 
The red curve represents the case with no restriction on the minimum volume of individual regions whereas the blue curve corresponds to one where the minimum allowed supercluster volume is $216 ~\mpc^3$. 
The black dashed vertical line represents the onset of percolation as explained in Figure \protect\ref{fig:FF_LCS}. 
Green dot on the blue curve shows the resulting sample we get by fixing the threshold at $\dth = 5.0$, which contains $N_{\rm R}=968$ superclusters with volume greater than $\vth \geq 216 ~\mpc ^3$.}
\label{fig:NC}
\end{figure}

\begin{figure*}
\centering
\subfigure[]{
\includegraphics[width=0.485\textwidth]{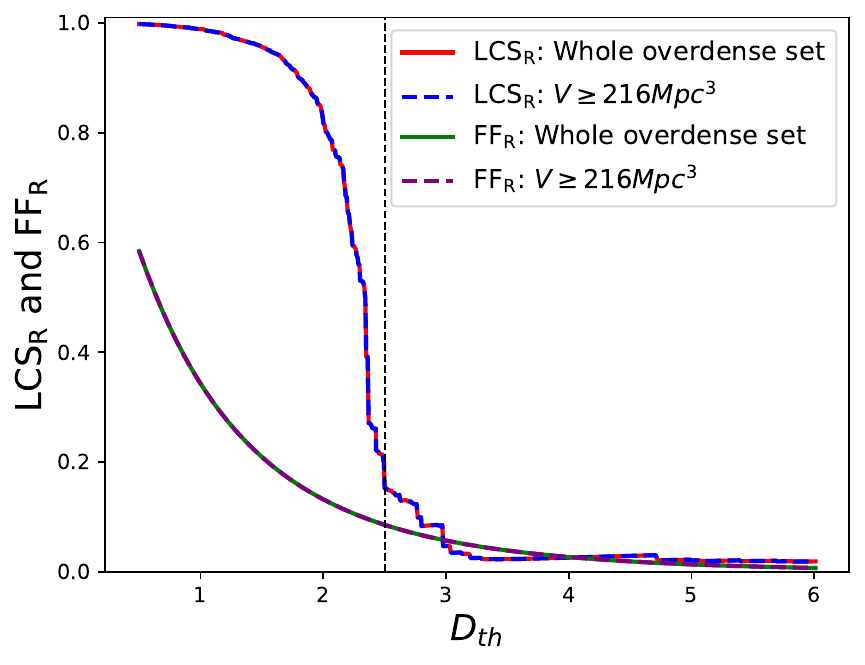}}
\subfigure[]{
\includegraphics[width=0.485\textwidth]{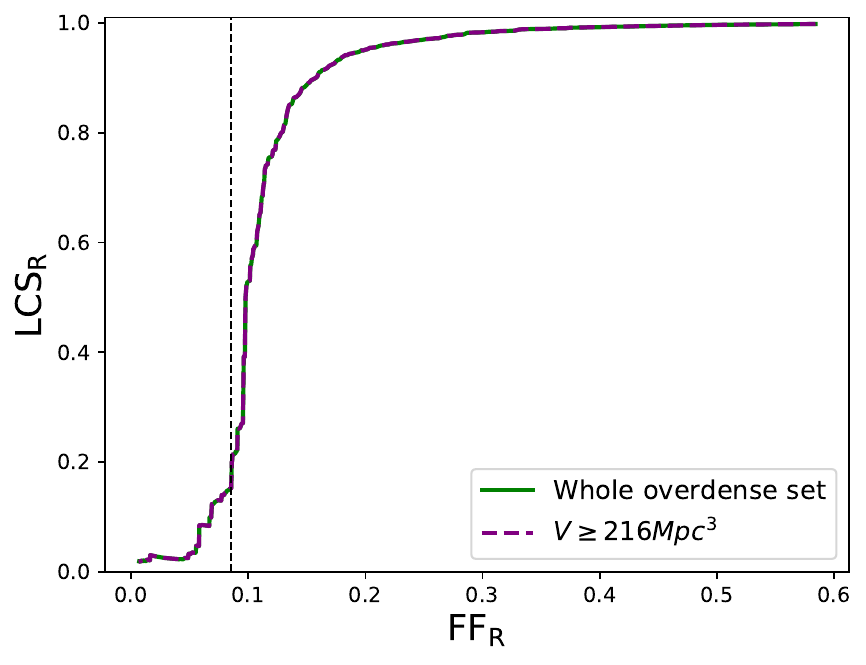}}
\caption{The left panel shows the filling factor (FF) and the largest cluster statistics (LCS) of the overdense volume as a function of the density threshold $D$. 
In the right panel LCS is shown as a function of FF. 
The solid curves in both the panels include all the regions whereas for the dashed curves we consider only the regions which have volume larger than a cutoff $\vth=216 ~\mpc ^3$.}
\label{fig:FF_LCS}
\end{figure*}

In Figure~\ref{fig:NC}, we plot \nr~ as a function of the density threshold $\dth$. The red curve shows all the isolated overdense regions. 
Naturally, an arbitrarily small isolated overdense region cannot be considered a supercluster. 
Therefore, in order to define superclusters, we first impose the requirement of a minimum size which is of the order of the smoothing scale (this condition is consistent with that used in \citet{Liivamagi:2010jg, Einasto:2011zc}). 
In terms of volume, the condition translates into requiring a minimum volume of $\vth \sim (a/2)^3=216~\mpc^3$, where $a$ is the scale of the smoothing kernel.
The blue curve in Figure~\ref{fig:NC} includes only the regions which have volume larger than the cutoff $216~\mpc^3$. 

Figure~\ref{fig:FF_LCS} shows the behaviour of the filling factor and largest cluster statistics.
In the left panel, we plot the FF and LCS against the density threshold, and the right panel illustrates the relation between FF and LCS. 
Percolation transition can be identified as a discontinuous change in LCS while plotted against filling factor \citep{Yess1996, Shandarin:1997fc}; see also \citet{J2018}. 
From Figure~\ref{fig:FF_LCS}, we also identify that the density field percolates at $D \lesssim 2.5$, which corresponds to the filling factor $FF \gtrsim 8.6 \%$. 
This density threshold value at the onset of percolation coincides with the corresponding threshold found by \citet{Liivamagi:2010jg} for the SDSS DR8 data.
The percolation transition is shown by the black dashed vertical lines in Figures~\ref{fig:NC} and \ref{fig:FF_LCS}. 
At the percolation transition, LCS rises abruptly as we decrease $\dth$ slightly while the filling factor increases slowly and smoothly.
Note that the number of the isolated regions ($N_R$) exhibits a peak (or a turn-around) in Figure~\ref{fig:NC} near the percolation transition for both with and without a minimum volume requirement. 
Percolation analysis not only helps us to understand how the connectedness of the field changes with the threshold but can also be useful to select the thresholds for defining \scls in both fixed and adaptive threshold approaches. 
The critical threshold at the onset of percolation transition is important for setting the minimum threshold while finding superclusters with individual thresholds found in an adaptive approach.

We also see that imposing supercluster volume cutoff does not affect volume fractions significantly. 
The difference in the number of objects is in the order of several hundreds -- almost a quarter, however the combined volume of those regions is negligible.

\subsection{Selecting threshold density to define  superclusters}
\citet{Einasto:2011zc} analysed how the size and richness of 
superclusters in the largest nearby 
supercluster system, the Sloan Great Wall, change with the change of the 
threshold density used to define superclusters. They showed that at lower
threshold density, $\dth=4.0$, the superclusters in the Sloan Great Wall are 
merged with the surrounding superclusters. As $\dth$ increases,
individual superclusters start to separate. At very high $\dth$,
only the highest density cores of superclusters remain.
Based on this analysis, \citet{Einasto:2011zc} suggested that the density threshold to define individual superclusters should be close to $\dth=5.0$.
Therefore, following \citet{Einasto:2011zc} we set $\dth=5.0$ (as shown by the green dashed vertical line in Figure \ref{fig:NC}) in order to define the superclusters for this fixed threshold approach. 
We obtain $N_{\rm R}=968$ in this case with the volume cutoff 
$\vth=216 ~\mpc ^3$ (shown by the green dot on the blue curve 
in Figure~\ref{fig:NC}). 
In Appendix \ref{app:Flat_diff_ths}, we also show the results of 
the analysis of supercluster morphology using two other density thresholds, $\dth=4.5$
and $\dth=5.5$.

\subsection{Finding galaxies within superclusters}
In the previous section, we studied superclusters as isolated regions with the density higher than a threshold ($\dth=5.0$) and volume larger than a volume cut off $\vth=216~\mpc^3$. 
Superclusters, however, are not just density enhancements, they are primarily collections of galaxies. 
From here onwards, we also place the galaxies inside the supercluster boundary surfaces we just determined. 
Additionally, we impose an extra constraint that a supercluster must have two or more galaxies in order to have an easy definition for the extent of the supercluster. 
This condition may reduce \scl numbers slightly but the density level for percolation remains almost unchanged.

To assign galaxies in superclusters, we use the coordinates of the galaxies observed in the SDSS DR12 main galaxy sample \citep{Liivamagi:2010jg} along with the groups found on the same dataset by \citet{Tempel:2017gr}. 
The procedure we follow is described below.

\begin{itemize}
    \item  For each galaxy we first find out if it falls into a supercluster or not. For this purpose we check if any of the eight vertices of the cell containing a galaxy falls inside a supercluster (recall that the volume of the supercluster has to be larger than $\vth=216~\mpc^3$). In such a case we assign the galaxy to that supercluster. 
    
    \item If a galaxy is part of a group, we keep track of all the member galaxies of that group. If any of the member galaxies falls into a supercluster we assign all the member galaxies of the group to that particular supercluster\footnote{Assuming that a group as a gravitationally bound object has to be either fully inside or fully outside a supercluster.}. 
    \item In unlikely occasions if a group is shared by two or more superclusters we assign the whole group to the supercluster having the maximum number of member galaxies.
    
    \item Once we complete assigning all the galaxies to the superclusters, we look for the superclusters having less than two galaxies, if any at all. If such objects exist, we identify them as spurious and discard them from the final supercluster catalogue.
\end{itemize}

\section{Minkowski functionals and Shapefinders}\label{sec:Minkowski}

The morphology of a closed two dimensional surface embedded in three dimensions is well described by the four Minkowski functionals (MFs) which are as follows \citep{mecke}
\begin{enumerate}
 \item Enclosed volume: $V$,
 
 \item Surface area: $S$,
 
 \item Integrated mean curvature (IMC):
 \begin{equation}
  C=\frac{1}{2} \oint \left(\frac{1}{R_1}+\frac{1}{R_2} \right) dS \;,
 \end{equation}
 
 \item Integrated Gaussian curvature or Euler characteristic: 
 \begin{equation}
 \chi=\frac{1}{2\pi} \oint \frac{1}{R_1 R_2} dS\;.
 \end{equation} 
 Here $R_1$ and $R_2$ are the two principal radii of curvature at any point on the surface. 
\end{enumerate}

The fourth MF (Euler characteristic) can be written in terms of the genus ($G$) of the surface as follows,
\begin{equation}
 G=1-\chi/2 \equiv {\rm (no.~ of~ tunnels)}-{\rm (no.~ of~ isolated~ surfaces)}+1\;.
\end{equation}
It is well known that $\chi$ (equivalently $G$) is a measure of the topology of the surface.
 For a closed isolated surface, the genus number is an integer that measures how many `tunnels' (or `holes') pass through the surface. In other words, an isolated surface with $n$ tunnels will have genus $G=n$ which is invariant under any continuous deformation of the surface. For example, a simply connected surface (such as a spherical surface or any deformation of it) has no tunnel or hole passing through it and its genus value is zero, $G=0$, that represents its `trivial topology'. On the other hand, a doughnut and a coffee mug, both have one tunnel passing through them, therefore both have $G=1$. 
We refer to \citet{Nakahara:2003nw} for a detailed review about the topology and genus and \citet{mecke} for a comprehensive discussion on applications of MFs including genus in cosmology.

The `Shapefinders', introduced in \citet{Sahni:1998cr}, are ratios of the MFs,
namely
\begin{align}
 {\rm Thickness:}~T &=3V/S\;, \nonumber \\
 {\rm Breadth:}~B &=S/C\;, \label{eq:L} \\
 {\rm Length: }~L&=C/(4\pi)\;. \nonumber
\end{align}
Shapefinders -- $T, B, L$ -- have dimension of length, and estimate the three physical extensions of an object in 3-dimensions\footnote{In general one finds $L \geq B \geq T$.
However, if the natural order $T\leqslant B \leqslant L$ is not maintained, we choose the smallest dimension as $T$ and the largest one as $L$ to restore the order. 
In rare cases where a cluster has $C < 0$ we shall redefine $C \to |C|$ to ensure that $B$ and $L$ are positive.}.
The Shapefinders are spherically normalized, i.e. $V=(4\pi/3) T B L$. 
 
Using the Shapefinders one can determine the morphology of an object (such as
an isodensity surface), by means of the following dimensionless quantities\footnote{One can redefine `length' by taking the genus of an object into account \citep{Sheth:2002rf},
\begin{equation}\label{eq:L1_HI}
L_1=\frac{C}{4 \pi (1+|G|)}\;. \nonumber
\end{equation}
 This reduces the filamentarity in the following manner while keeping planarity unchanged,
 \begin{equation}\label{eq:f1_HI}
  F_1=\frac{L_1-B}{L_1+B}\;. \nonumber
 \end{equation}
 These definitions, $L_1$ and $F_1$, receptively assess the `macroscopic' length and filamentarity of a given object. 
 On the other hand, the definitions of length and filamentarity, given in Eq. \eqref{eq:L} and \eqref{eq:F} respectively, provide us with the `microscopic' information which we are interested in this work.}
 which characterize its planarity and filamentarity \citep{Sahni:1998cr}
\begin{align}
{\rm Planarity:}~ P&=\frac{B-T}{B+T}\;,\label{eq:P} ~~\\ 
{\rm Filamentarity:}~ F&=\frac{L-B}{L+B}\;. \label{eq:F}
\end{align}

 A simply connected spherical and a planar surface both can have same trivial topology with genus zero. However, their morphology can be easily distinguished in terms in terms of the Shapefinders, especially by the quantities -- $\lbrace P, F \rbrace$ (both are in the range $\in (0,1)$).
For a planar object (such as a sheet) $P \gg F$, while the reverse is true for a filament which has $F \gg P$. 
A ribbon will have $P \sim F$ whereas $P \simeq F \simeq 0$ for a sphere.
In all cases $ 0 \leq P,F \leq 1$. 
Therefore, Shapefinders, together with MFs, provide us with all the information about the geometry, morphology and topology of a 3-dimensional field.

\subsection{SURFGEN2 algorithm}
To quantify the shape of the superclusters we employ SURFGEN2, which is an advanced version of the SURFGEN algorithm originally developed by \cite{Sheth:2002rf, Sheth:2006qz}. The details of the refinements in SURFGEN2 can be found in \citet{Bag:2018fyr,Bag:2018zon}. 
However, for completeness, here we also briefly discuss the SURFGEN2 code which constructs isodensity surfaces from a given density field and subsequently calculates their Minkowski functionals and Shapefinders.

\begin{itemize}
 \item SURFGEN2 code first identifies all the isolated regions above or below a given threshold in the field (e.g. a density field) within the sample volume using the friends-of-friend algorithm. 
 Since we study galaxy superclusters in this work, we focus on the overdense part. SURFGEN2 can also find the regions consistent with periodic boundary conditions; however, it is not required in this particular work.

 \item Next, the code models the surface of each supercluster by triangulating the surfaces. 
 The classic SURFGEN \citep{Sheth:2002rf, Sheth:2006qz} triangulates the surfaces using the Marching Cube algorithm \citep{marcube}, which has a few issues like hole formation, degeneracy etc. 
 SURFGEN2, however, uses an improved triangulation scheme, known as `{\em Marching Cube 33}' \cite{mar33}, which circumvents the issues associated with the classic Marching Cube algorithm.
 
 \item As the last step, the MFs (and Shapefinders) are calculated for each individual supercluster using the stored triangle vertices. 
\end{itemize}

Note that one could also focus on the under-dense regions in the density field and study those in the same manner.

\section{Studying supercluster morphology using Shapefinders}

\subsection{Superclusters obtained using a fixed density threshold}\label{sec:fixed_dth}

In this approach we define the superclusters as isolated regions which obey the following conditions.
\begin{enumerate}
    \item Density has to be higher than a constant fixed threshold, i.e.,
    \beq
    D({\bf r}) \geq \dthc \;,
    \eeq
    everywhere inside the region.
    \item Volume of each region has to be greater than the cutoff that corresponds to a minimum size of the order of the smoothing scale, i.e.,
    \beq
    V \geq \vth = (a/2)^3=(6 ~\mpc)^3 = 216 ~\mpc ^3 \;,
    \eeq
    where $a=12 ~\mpc$ is our smoothing scale.
    \item Must include two or more galaxies.
\end{enumerate}

\begin{figure}
\centering
\includegraphics[width=\linewidth]{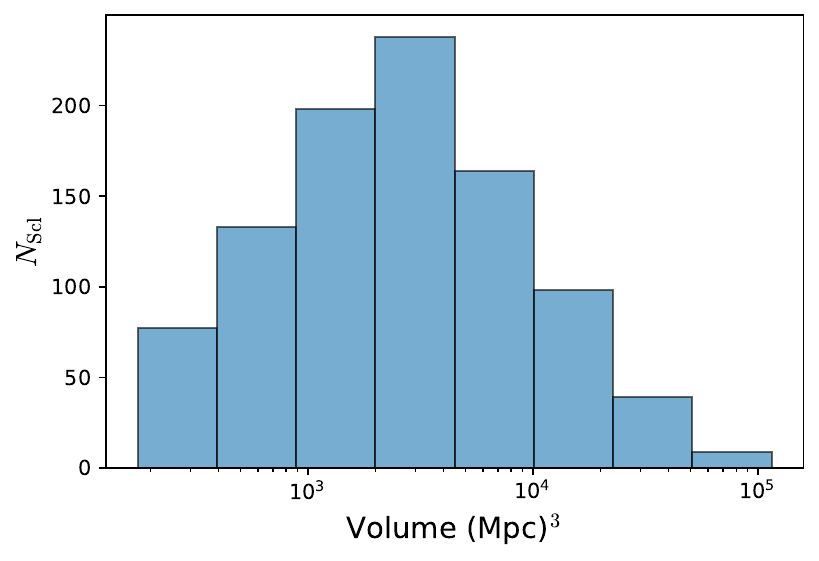}
\caption{Volume distribution of \scls for the fixed $\dth=5.0$ sample. The distribution peaks at $\sim \mathcal{O}(10^3 \mpc^3)$. }
\label{fig:Flat_vol_dist_Dth5}
\end{figure}

Following \citet{Juhan_thesis,Einasto:2011zc, Liivamagi:2010jg}, we choose $\dthc=5.0$ obtaining \nscl$=956$ superclusters. The volume distribution is portrayed in Figure~\ref{fig:Flat_vol_dist_Dth5}. We find most of the \scls in the volume range $10^3 - 10^4 ~\mpc^3$.
We show the statistics of supercluster samples corresponding to a few different thresholds in Table~\ref{tab:flat}, where our primary choice $\dthc=5.0$ is highlighted in boldface. 
One can notice that, as we increase the density threshold, the quantities like \nc, filling factor, fraction of galaxies inside \scls decrease. 
In Appendix \ref{app:Flat_diff_ths}, we demonstrate that large \scls are fewer, but they dominate the total supercluster-volume. This behaviour stands valid for all thresholds. 

\begin{center}
\begin{table*}
\renewcommand{\arraystretch}{1.8}
 \begin{tabular}{||c|c||c|c|c|c|c|c||} 
 \hline
 $D_{th}$ & \makecell{No. of \\ isolated \\ overdense \\ regions (\nr)} & \makecell{No. of \\ superclusters \\ (\nscl)}  & \makecell{Total no. of \\ grid cells inside \\ \scls \\ (\ncell)} & \makecell{Total volume \\estimation from \\ SURFGEN2 \\ $(\mpc^3)$}  & FF $(\%)$ & \makecell{No. of galaxies \\ inside \\ \scls \\ (\ng)} & \makecell{Fraction of \\ galaxies inside \\\scls \\ ($f_{\rm gal}~{\rm in}~\%$)} \\ [0.5ex] 
 \hline\hline
 $4.5$ &  $ 1143 $ &  $ 1127 $ & $ 7907370 $ & $ 7852772.11$ &  $ 1.89 $ & $  105111$ & $ 17.985$    \\ 
 \hline
 
 $4.9  $ &  $ 998 $ &  $988  $ & $ 6014295 $ & $5969329.27 $ &  $  1.44$ & $ 87785 $ & $15.020 $    \\ 
 \hline
 
 $ {\bf 5.0} $ &  $ {\bf 968}$ &  $ {\bf 956} $ & $  {\bf 5626042}$ & $ {\bf 5583108.53}$ &  $  {\bf 1.34}$ & ${\bf  83815} $ & $ {\bf 14.34}$    \\ 
 \hline
 
 $ 5.1 $ &  $  939$ &  $  928$ & $ 5264661 $ & $ 5223605.97$ &  $  1.26$ & $80422  $ & $ 13.76$    \\ 
 \hline
 
 $ 5.5 $ &  $ 826  $ &  $  823$ & $  4044665$ & $ 4010992.76$ &  $ 0.97 $ & $ 67786 $ & $11.60 $    \\ 
 \hline

 \hline
\end{tabular}
\caption{Statistics of supercluster samples extracted using different fixed thresholds. The minimum volume is $\vth=216 ~\mpc^3$ in all cases and, furthermore, isolated overdense regions have to include at least two galaxies (causing a slight discrepancy between \nscl~ and \nr). Notice that counting grid cells inside a supercluster is a good estimate of its volume (as long as the volume is quite large). As expected, the filling factor and the fraction of galaxies inside \scls decrease as we increase the $\dth$. }
\label{tab:flat}
\end{table*}
\end{center}

\begin{figure*}
\centering
\hspace*{-1mm}
\subfigure[Scl No. 725: small (spherical); \protect\newline ~~~$V=295.93~\mpc^3, P=0.01, F=0.02$;
\protect\newline ~~~ $N_{\rm grp}=8, N_{\rm gal}=57$;  \protect\newline ~~~ Origin: (RA, dec, $z$) = ($234.96, 30.72, 0.10$).]{
\includegraphics[width=0.33\textwidth]{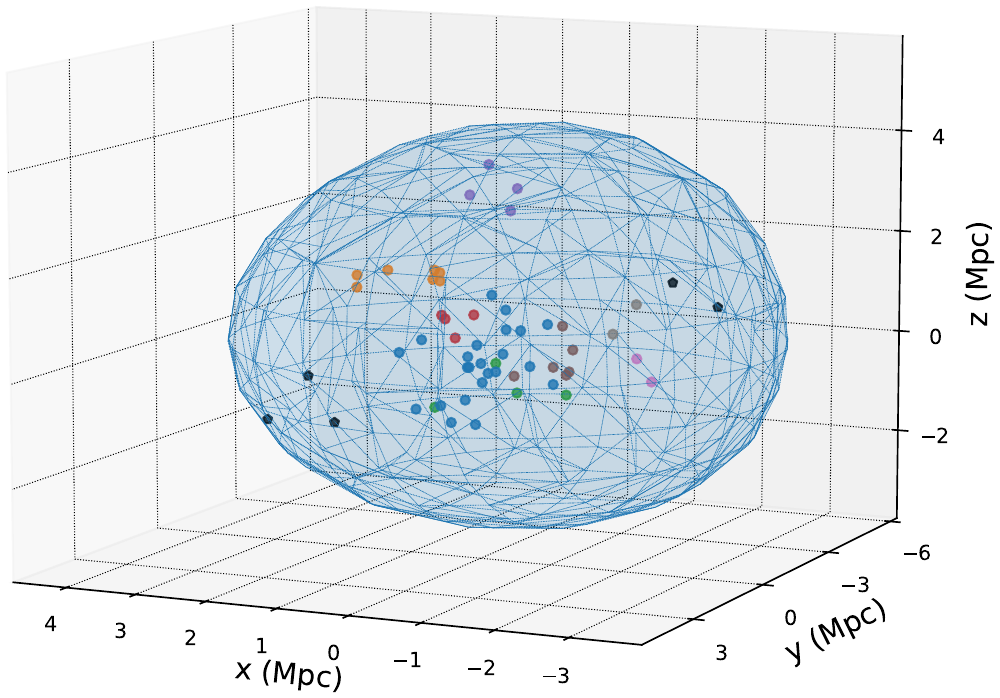}}\hspace*{-1mm}
\subfigure[Scl No. 548: elongated; \protect\newline ~~~ $V=2.80 \times 10^{4}~\mpc^3, P=0.06, F=0.34$;
\protect\newline ~~~ $N_{\rm grp}=33, N_{\rm gal}=182$; \protect\newline ~~~ Origin: (RA, dec, $z$) = ($211.86, 27.24, 0.16$).]{
\includegraphics[width=0.33\textwidth]{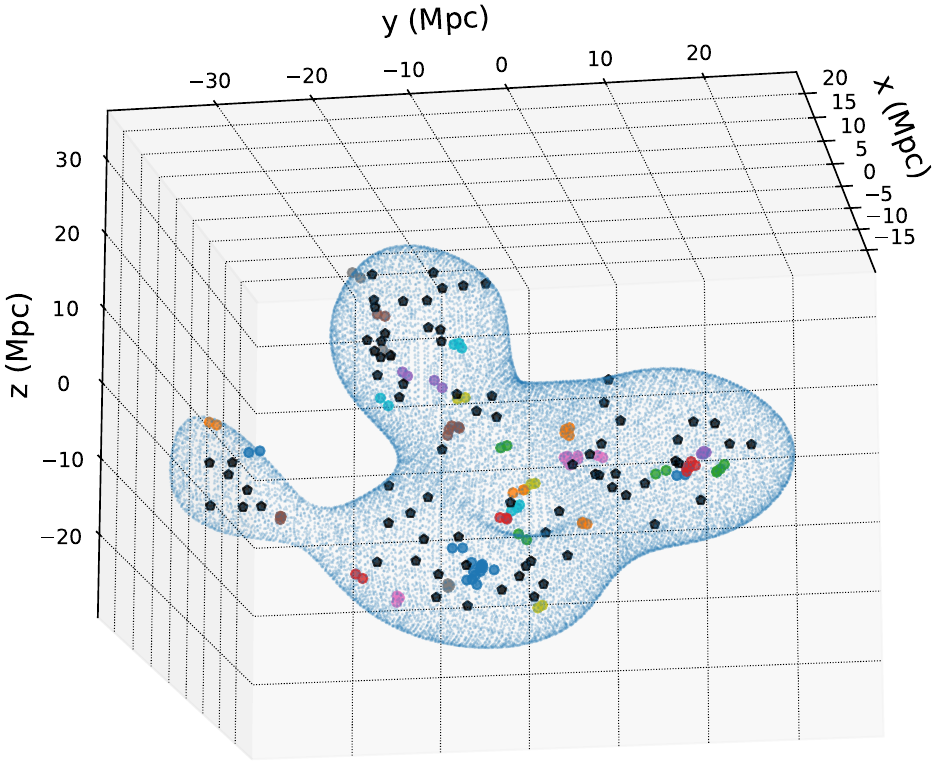}}\hspace*{-1mm}
\subfigure[Scl No. 197: large (filamentary); \protect\newline ~~~ $V=1.15 \times 10^{5}~\mpc^3, P=0.11, F=0.69$;
\protect\newline ~~~ $N_{\rm grp}=741, N_{\rm gal}=4901$; \protect\newline ~~~ Origin: (RA, dec, $z$) = ($184.42, 3.67, 0.08$).]{
\includegraphics[width=0.33\textwidth]{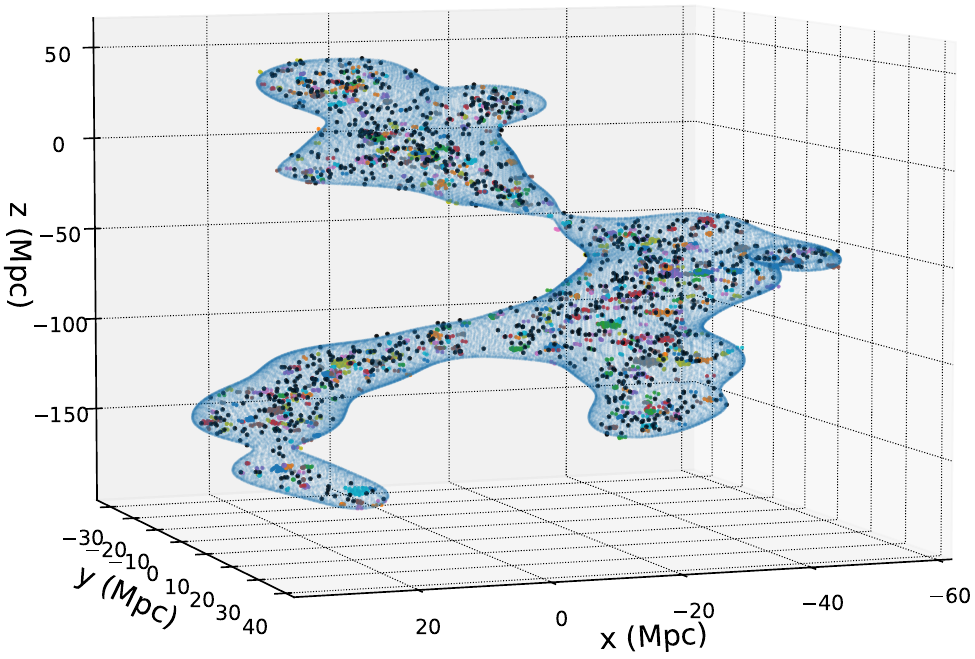}}
\caption{The surfaces of three \scls -- a small (left panel), a moderately large (middle panel) and the largest supercluster (right panel) -- from our fixed $\dth=5.0$ catalog have been shown in 3-dimensions for demonstration. Note that only for the small \scl in the left panel, we show the triangulated surface as modelled by SURFGEN2 whereas for the other two relatively large \scls we depict the surface by plotting the midpoints of the triangles as the triangles are too many in number and too small in size. All three \scls have $G=0$. We also plot the galaxy distribution inside these superclusters. Galaxies belonging to different groups are marked by filled circles with different colors whereas the lone galaxies that are not member of any group are shown by black pentagons. The origin of each plot has been shifted to the richest group inside the supercluster; the RA, dec and $z$ of the richest group is mentioned in the respective panel caption.
The panel captions also mention the size, shape of the respective \scls as well as how many groups and galaxies they enclose. Note that Scl No. 197 given in the right most panel is identified as the richest supercluster in the Sloan Great Wall.   
}
\label{fig:vis_scls_g0}
\end{figure*}

Since the MFs (strictly speaking the first three, $\lbrace V,S,C \rbrace$) and the Shapefinders  (SFs: `thickness', `breadth' and `length' defined in \eqref{eq:L}) are likely to increase with the volume of the \scls by definitions, it would be less interesting to study them for individual superclusters. 
Hence, in this work we primarily focus on the morphological parameters, `planarity' ($P$) and `filamentarity' ($F$), and the topology, described by genus ($G$), of each individual \scl (although we calculate the previously mentioned geometrical quantities in the intermediate steps).

In Figure \ref{fig:vis_scls_g0} we show the boundary surfaces of three superclusters, as modelled by the Marching Cube 33 triangulation using SURFGEN2, in 3-dimensions along with the galaxy distribution in them. For this illustration, we chose three \scls with trivial topology ($G=0$) as examples -- a small one, a moderately large one and the largest \scl in our galaxy sample, arranged from left to right. We adopt the same Cartesian grid as in our density field but the origin in each plot has been shifted to the location of the richest galaxy group (that includes maximum number of galaxies) inside the respective supercluster. The RA, dec and $z$ of the richest group is given in the respective panel caption which also mentions the volume ($V$), planarity ($P$) and filamentarity ($F$) of the supercluster together with how many groups and galaxies it encloses. For the smallest supercluster, we depict the triangulated surface whereas for the larger ones we show the surface by plotting the centre of the triangles for better visualisation as there are too many triangles. We also show the galaxy distribution in each supercluster; member galaxies of different groups are shown by different coloured circles and the lone galaxies that do not belong to any group are shown by the black pentagons. It is visually evident that the small \scl (left panel) has quite a spherical shape that is supported by our findings of low planarity and filamentarity for this supercluster: $P \sim 0$ and $F \sim 0$. The moderately large \scl in the middle panel can be seen quite elongated and its filamentarity attains an intermediate value of $F \approx 0.34$ while the planarity remains low, $P \approx 0.06$. The largest \scl (right panel) is extremely galaxy rich as it contains $741$ galaxy groups and $4901$ galaxies in total. In fact, we identify this supercluster (Scl No. 197)
as the richest supercluster in the Sloan Great Wall (SGW) in our catalog (with fixed $\dth=5.0$)\footnote{When we increase the flat threshold to $\dth=5.5$, we find that Scl No. 197 breaks into several branches in our catalog.}. 
It is visually quite filamentary, which reflected in it's high filamentarity value, $F \approx 0.69$ and low planarity, $P\approx 0.11$. In all these cases, the Shapefinders, calculated using SURFGEN2, support the visual assessment.

\subsection{Results with constant threshold}

\begin{figure}
\centering
\includegraphics[width=\linewidth]{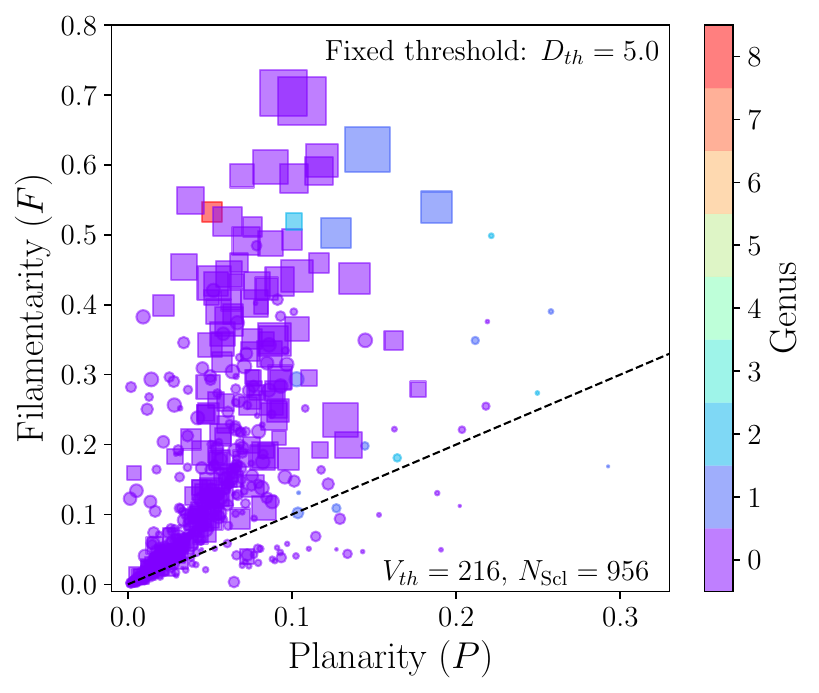}
\caption{Distribution of Shapefinders in the fixed threshold supercluster sample. The smaller ($V < 10^{4}$ Mpc$^3$) and larger ($V \geq 10^{4}$ Mpc$^3$) \scls are marked by the filled circles and squares respectively where their volumes are proportional to the size of the markers. The genus value of the \scls is indicated by the colour bar. The dashed line represents $F=P$ straight line above which most \scls lie.}
\label{fig:Flat_SF_Dth5}
\end{figure}

 The planarity ($P$) and filamentarity ($F$)  of individual \scls corresponding to $\dthc=5.0$ are shown in Figure~\ref{fig:Flat_SF_Dth5} by scattered filled circles and squares for smaller ($V < 10^{4}$ Mpc$^3$) and larger ($V \geq 10^{4}$ Mpc$^3$) \scls respectively. The limit, $V=10^4~ \mpc^3$, here is roughly the same limit found in \citet{M2011c} that separates large elongated \scls from the small spherical ones. 
The size of the markers (circles and squares) is proportional to the volume of the superclusters. The dashed line shows the $F=P$ straight line above which most of the \scls lie.
Genus values of the \scls are shown by different colours. 
We see that smaller \scls are more spherical, having both $P$ and $F$ very small.
The larger \scls are more filamentary (most large \scls have $F>P$).  Although some large \scls have slightly complex topology (genus $G > 0$), 
most superclusters, however, have trivial topology with zero genus.
Interestingly, a few outlier small \scls are found to be quite planar or filamentary with $G>0$.
Figure~\ref{fig:Flat_SF_Dths} in Appendix~\ref{app:flat_SF_dths} shows $P$ and $F$ of individual \scls for two more thresholds $\dth=\{4.5,5.5\}$ for comparison.

\begin{figure*}
\centering
\subfigure[]{
\includegraphics[width=0.465\textwidth]{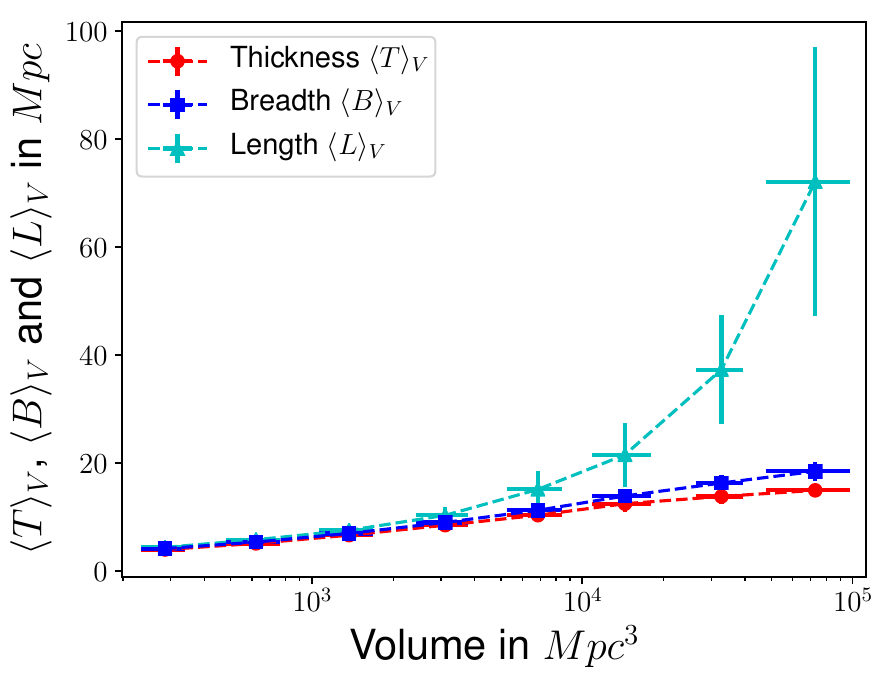}}
\subfigure[]{
\includegraphics[width=0.508\textwidth]{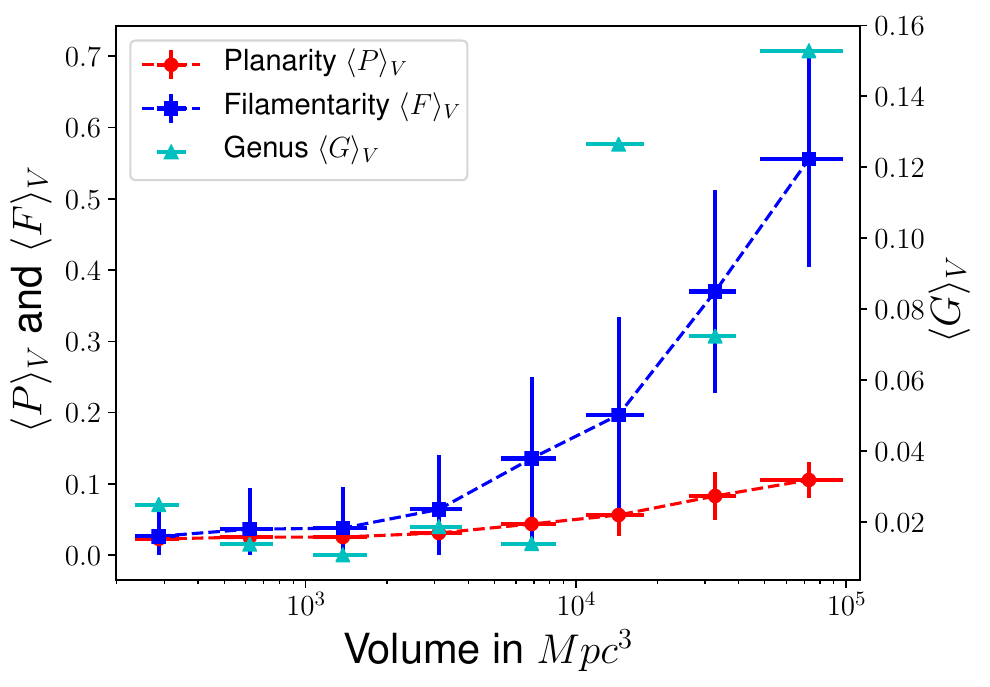}}
\caption{Shapefinders of the \scls are plotted against their volume, divided into $8$ bins, in the fixed threshold $\dth=5.0$ case. 
For each volume bin we plot the volume-averaged values of the Shapefinders: namely $\langle T \rangle_V, \langle B \rangle_V$, and $\langle L \rangle_V$ in the left panel and $\langle P \rangle_V, \langle F \rangle_V$, and $\langle G \rangle_V$ in the right panel. 
Error bars represent the standard deviations which show the scatter of the respective quantities in each bin (except for the genus). 
}
\label{fig:Flat_SF_Dth5_v}
\end{figure*}

In Figure~\ref{fig:Flat_SF_Dth5_v}, we plot the Shapefinders against the volume of superclusters, divided into 8 bins. We show the volume-averaged Shapefinders in each bin (i.e. average of Shapefinders of \scls in each bin weighted by their volumes). 
Error bars represent the standard deviation (scatter) of the respective quantities in each bin. We clearly see that smaller \scls are mostly spherical with $T \approx B \approx L$ (therefore, $P \approx 0 \approx F$). 
On the other hand, as we move to the larger volume bins we find from the left panel that `length' rises much more rapidly than the other two Shapefinders, $L \gg T \approx B$. 
This consequently results in steeper rise in the `filamentarity' as compared to the `planarity', as evident in the right panel. Therefore, the larger \scls are quite filamentary ($L \gg T \approx B$ which leads to $F > P$). 
We also notice that larger \scls tend to be slightly more multiply connected compared to the smaller ones as demonstrated by the (volume-averaged) genus curve in the right panel.

 As mentioned above in the end of the Section \ref{sec:fixed_dth}, the largest supercluster in our catalog (with fixed $\dth=5.0$), Scl No. 197 (shown in the rightmost panel of Figure \ref{fig:vis_scls_g0}), corresponds to the Sloan Great Wall. Similarly we identify the Scl No. 651 as another famous supercluster, the Corona Borealis supercluster. We find that both these large superclusters are quite filamentary with $(P,F)=(0.11, 0.69)$ and $(0.10,0.58)$ for the Scl No. 197 and 651 respectively. These numbers, as well as our overall assessments of the shape of the SGW and CB superclusters (using the SURFGEN2 algorithm) are consistent with the previous findings in the literature \citep{Einasto:2011zc} based on SDSS DR7 data.
The slight difference in the estimations of filamentarity between our and  the analysis of \citet{Einasto:2011zc} can be arising from several factors, e.g. the adoption of larger smoothing scale and more refined SURFGEN2 algorithm in our work.

\begin{figure}
\centering
\includegraphics[width=\linewidth]{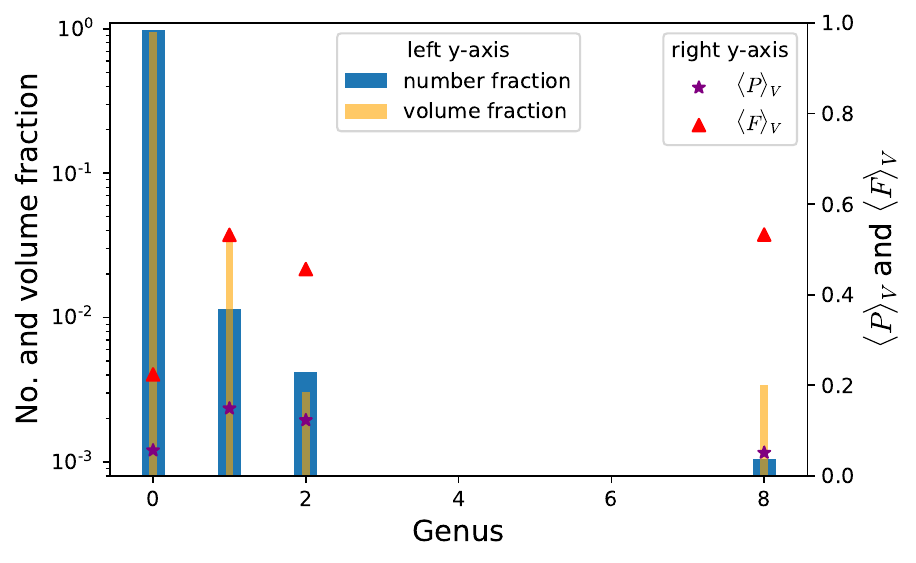}
\caption{Genus distribution in the constant threshold $\dth=5$ \scl sample. 
Blue and orange bars represent the fraction of \scls having different genus and the fraction of volume enclosed by those respectively (along left y-axis).
Purple stars and red triangles, respectively, show the volume-averaged planarity and filamentarity of the \scls for different genus values (plotted along the right y-axis).}
\label{fig:Flat_genus}
\end{figure}

\begin{figure*}
\centering
\subfigure[]{
\includegraphics[width=0.485\textwidth]{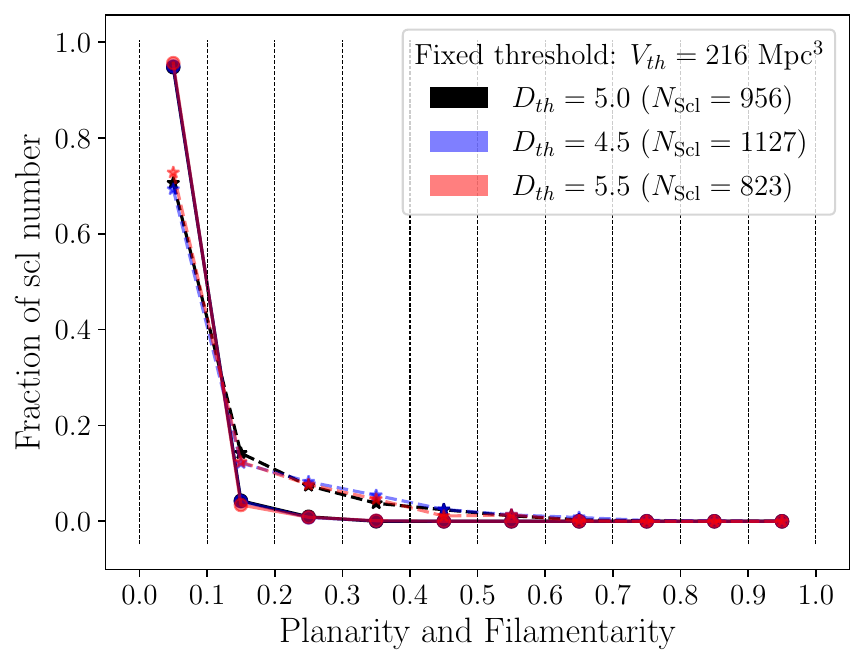}}
\subfigure[]{
\includegraphics[width=0.485\textwidth]{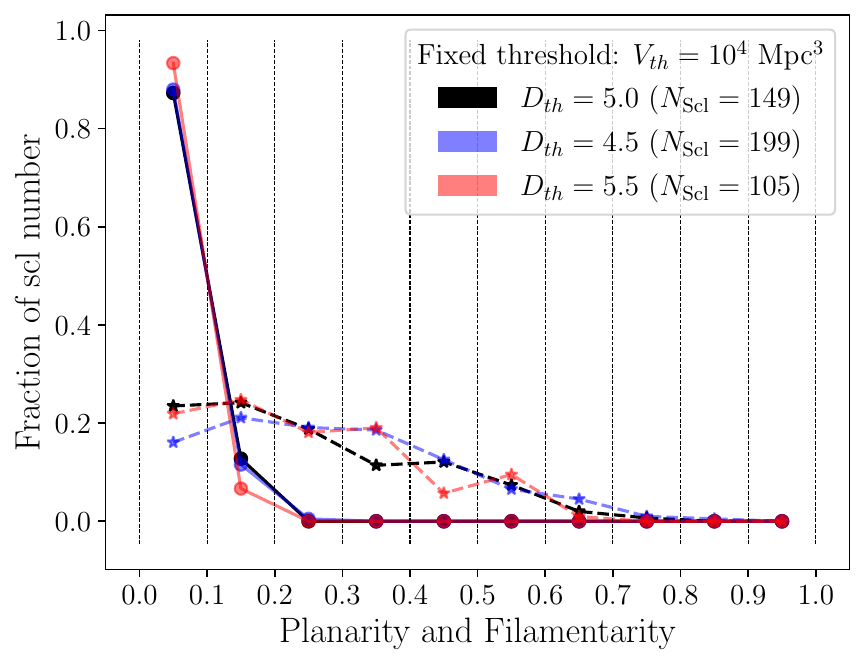}}
\caption{Distribution of the planarity (circles joined by solid lines) and the filamentarity (stars joined by dashed lines) of the \scl samples extracted using different fixed density thresholds.
The planarity and filamentarity ranges are divided into 10 bins.
Different colours represent the results for different fixed thresholds.
In the left panel we include all the \scls whereas in the right panel we focus on the large \scls with volume $\geq 10^4~\mpc^3$.
}
\label{fig:Flat_scl_number_dist}
\end{figure*}

\begin{figure*}
\centering
\subfigure[]{
\includegraphics[width=0.485\textwidth]{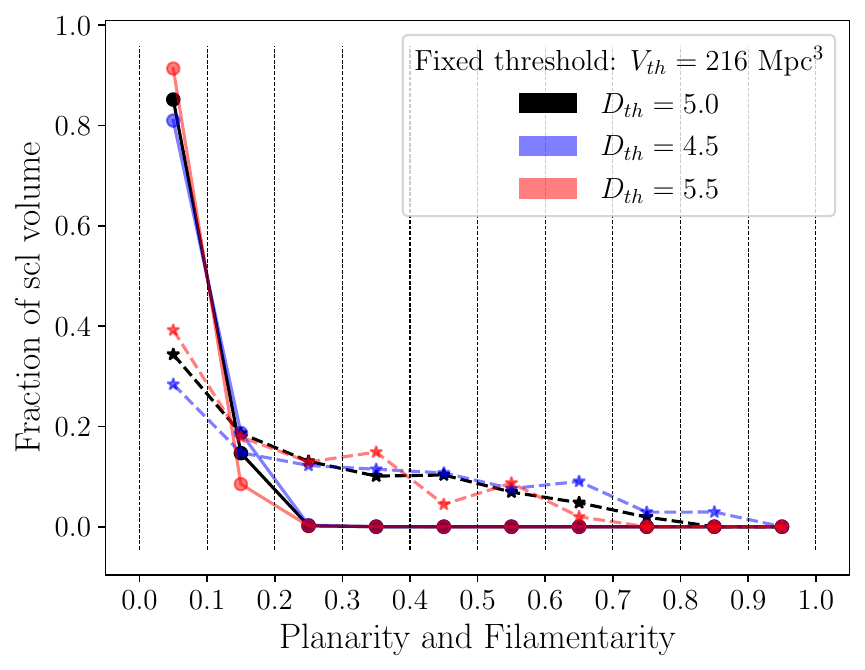}}
\subfigure[]{
\includegraphics[width=0.485\textwidth]{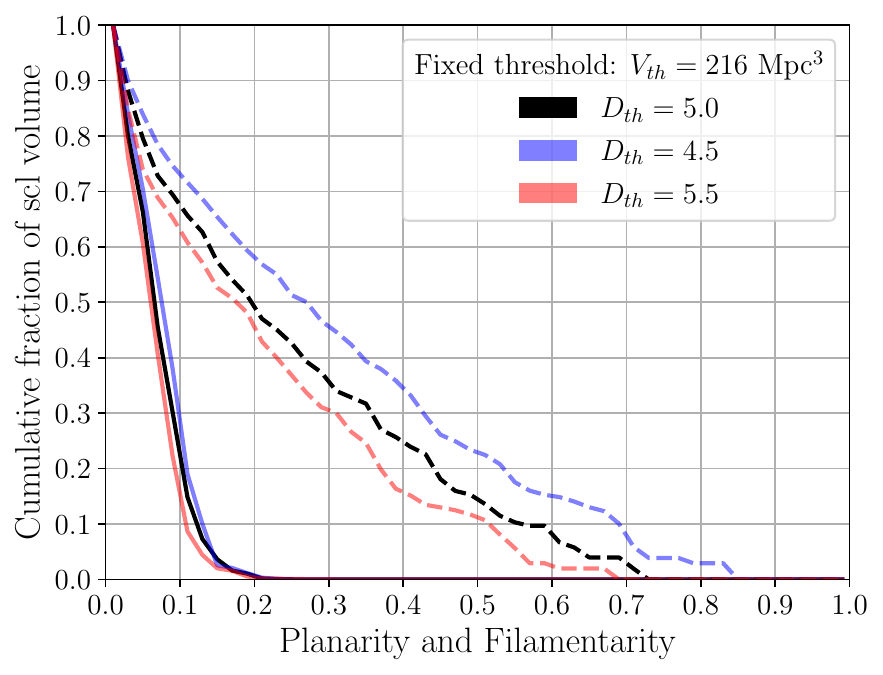}}
\subfigure[]{
\includegraphics[width=0.485\textwidth]{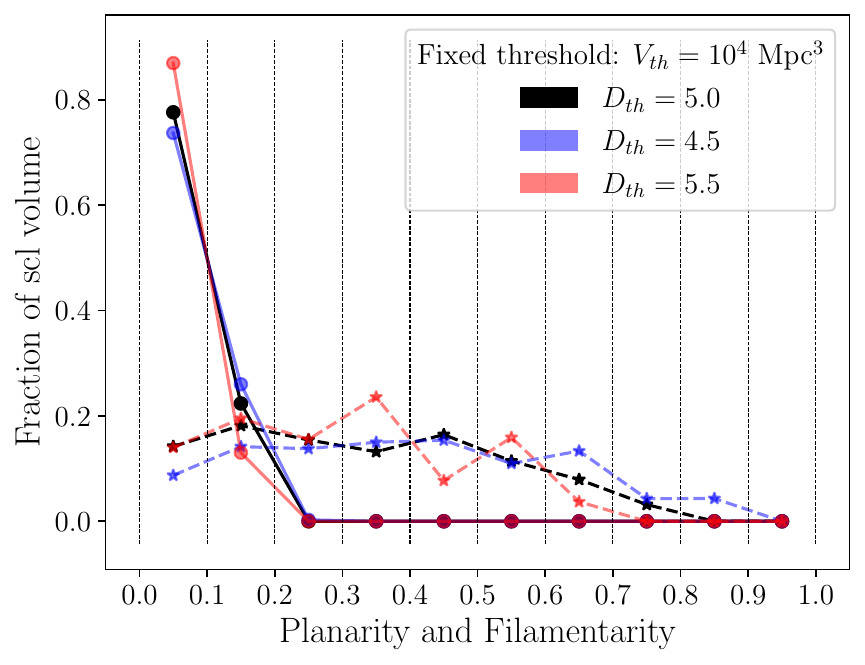}}
\subfigure[]{
\includegraphics[width=0.485\textwidth]{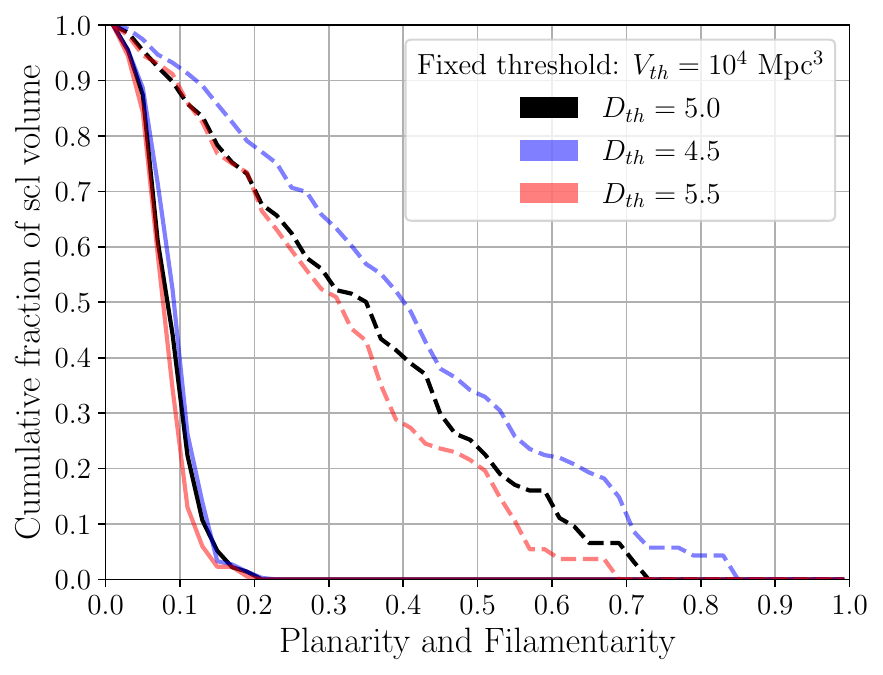}}
\caption{Volume weighted distributions of planarity (circles joined by solid lines) and filamentarity (stars joined by dashed lines) is depicted for \scl samples defined using various fixed thresholds (shown by different colours). Plots in the top row contain the full \scl samples whereas the bottom row panels include the large \scls with $\vth \geq 10^4 \mpc$ only.
Panels on the left, ({\bf a}) and ({\bf c}), show fraction of total supercluster-volume against the `planarity' ($P$) and `filamentarity' ($F$), both divided into $10$ bins. 
In the right panels, ({\bf b}) and ({\bf d}), we show the cumulative fraction of supercluster-volume in different planarity and filamentarity bins. 
Different colours represent results for different fixed density thresholds.}
\label{fig:Flat_frac_SF}
\end{figure*}

The statistics of \scls having different topology shown in Figure~\ref{fig:Flat_genus}. 
On the x-axis we plot the genus value, i.e. the number of tunnels passing through each supercluster.
The blue and orange bars respectively show the fraction of \scls having a specific genus and the fraction of supercluster-volume enclosed by them.
One can notice that almost $\sim 98\%$ of \scls have trivial topology, $G=0$, and these \scls account for roughly $ 95\%$ of the total supercluster-volume. 
Only a small portion of \scls (just $\sim 2\%$) have non-trivial topology, $G \geq 1$, and none have $G>2$ except the one with $G=8$.

Also in Figure~\ref{fig:Flat_genus} we show the volume-averaged planarity and filamentarity, plotted along right y-axis with purple stars and red triangles respectively, for \scls with different genus.
As genus increases the volume-averaged planarity remains small, $\langle P \rangle_V <0.2$ but we notice a significant increase in $\langle F \rangle_V$ among the \scls with non-trivial topology. 
This clearly demonstrates that high genus \scls tend to be significantly more filamentary. 

Figure~\ref{fig:Flat_scl_number_dist} shows the distribution of the shape/morphology parameters (planarity and filamentarity) of the \scls for the three values of the threshold (the primary choice of threshold is $\dth=5$, the other two are shown here for comparison only). 
The panel on the left shows the fraction of \scls in planarity and filamentarity bins. In the right panel we plot the same but only for the large \scls which have volume $\geq 10^4 ~\mpc^3$. 
Remarkably, for different thresholds we might have different \nc~(and filling factors) but the shape distributions are almost identical (especially in the left panel). 
We notice that \scls are statistically not planar at all, more than $90\%$ \scls have planarity less than $0.1$ in both panels.
As for filamentarity, similarly, the vast majority of the objects has very small values, meaning that most have a spherical shape.

However, larger \scls are significantly more filamentary as evident from the right panel -- only around $20 \%$ of these large \scls have $F < 0.1$ leaving roughly $80\%$ with $F > 0.1$.

Since larger \scls tend to be more filamentary, one may prefer to study the volume weighted shape distribution.
In the left panels of Figure~ \ref{fig:Flat_frac_SF}, we show the fraction of supercluster volume, defined as
\begin{equation}\label{eq:wV_shape}
   \frac{\sum ({\rm Volume ~of ~superclusters~falling ~in ~a~}P~{\rm or}~F ~{\rm bin})}{\rm Total ~volume ~of ~all ~the ~superclusters},
\end{equation}
falling into different planarity and filamentarity bins, again for the three density threshold values. 
Right panels show cumulative fraction, i.e. the fraction of the total supercluster-volume with planarity $P > P'$ is plotted against $P'$ (same goes for the filamentarity). 
We see that most of the supercluster volume is highly spherical in morphology (having both planarity and filamentarity very low) despite non-spherical superclusters typically being quite large.
Considering all the superclusters, (top panels ({\bf a}) and ({\bf b}) in Figure~\ref{fig:Flat_frac_SF}) we find that more than $85 \%$ of the supercluster-volume has $P<0.1$ while only around $30 -40 \%$ (depending on the threshold) of supercluster-volume has $F<0.1$ (despite having more than $70 \%$ of \scls falling in that filamentarity bin as seen in Figure~\ref{fig:Flat_scl_number_dist}).
For $\dth=5.0$, $15 \%$ of supercluster-volume has $F>0.5$. 
Focusing on the large \scls with $V \geq 10^4 \mpc^3$ (in bottom panels {\bf c}) and ({\bf d}), we again find that most of the supercluster-volume has very low planarity but now around $90 \%$ ($25 \%$) of volume of the large \scls has $F>0.1$ ($F>0.5$). 
In contrast, the planarity distribution is almost identical across the thresholds but at a lower fixed threshold slightly larger fraction of supercluster-volume is filamentary.

\begin{figure}
\centering
\includegraphics[width=\linewidth]{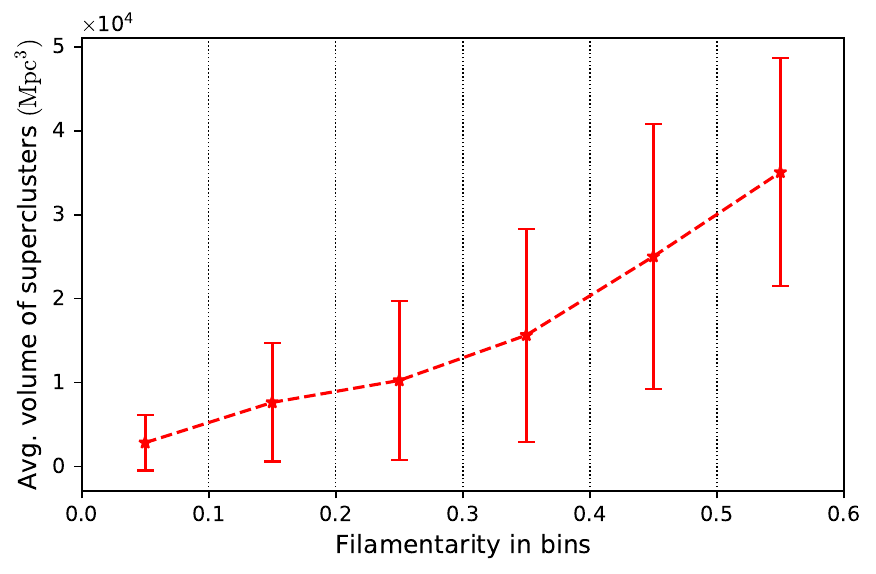}
\caption{Average volume of the \scls belonging to different filamentarity bins has been shown for the fixed threshold case ($\dth=5.0$). The error bars represent the standard deviation in each bin.}
\label{fig:Flat_F_bins}
\end{figure}

Since we find a moderate variation of filamentarity (and not planarity) among the superclusters, it would be interesting to see the volume distribution of \scls with different filamentarity. 
Figure~\ref{fig:Flat_F_bins} shows the average volume (along with the standard deviation) of the \scls belonging to different filamentarity bins for the fixed threshold $\dth=5.0$. 
It can be clearly seen that higher filamentary \scls are statistically larger in size.
This observation is consistent with that of Figure~\ref{fig:Flat_genus} and Figure~\ref{fig:Dths} explained in Appendix~\ref{app:Flat_diff_ths}.

Size distribution of density field objects acquired with this approach extends to smaller dimensions. It is worth mentioning that one may doubt whether the smaller (volume) `superclusters' are indeed astrophysical superclusters; some of them might be small structures like more isolated clusters or just random spurious density enhancements (especially on larger distances where number density of galaxies is lower). However, Appendix \ref{app:distance} demonstrates that the distribution of the geometrical properties, such as volume, weighted luminosity, morphology etc, of the \scls remain consistent across the Universe. Therefore, these somewhat suspicious small \scls are also homogeneously distributed in the Universe and hence do not bias our shape distribution results.


\subsection{Adaptive density threshold case}\label{sec:adaptive_method}
\begin{table}
\renewcommand{\arraystretch}{1.8}
\centering
 \resizebox{\linewidth}{!}{\begin{tabular}{||c|c|c|c||c|c|c||} 
 \hline
 $\dthmin$ &$\dthmax$ & $\vth$ ($\mpc^3$) & \nr & \nscl &  FF ($\%$) & $f_{\rm gal}$ ($\%$)\\ [0.5ex] 
 \hline\hline
 $2.5$ & $5.5$ & $1728$ & $1987$ & $ 1984 $ & $4.35 $ & $ 31.14$ \\ 
 \hline
 
 $2.5$ & $6.0 $ & $1728$ & $ 2005$ & $ 2002$ & $4.26 $ & $ 30.58$ \\ 
 \hline
 
 $2.5$ & $5.5 $ & $216$ & $3249 $ & $ 3201$   & $ 3.85$ & $ 30.51$  \\ 
 \hline
 \hline
 
  $3.0$ & $5.5 $ & $1728$ & $1589 $  & $1587 $  & $ 3.49$ & $26.80 $  \\ 
 \hline
 
 $3.0  $ & $ 6.0$ & $ 1728 $ & $1607 $ & $ 1605$   & $3.40 $ & $26.25 $  \\ 
 \hline
 
  $  3.0$ & $  5.5$ & $ 216 $ & $2639 $ & $2602 $   & $ 3.16$ & $26.73 $  \\ 
 \hline
 \hline
  ${\bf  3.5 }$ & $ {\bf  5.5} $ & ${\bf  1728 }$ & ${\bf  1271}$ & $ {\bf 1271}$   & ${\bf  2.76} $ & ${\bf  22.56}$  \\ 
 \hline
 
  $ 3.5 $ & $  6.0$ & $1728  $ & $ 1289$ & $ 1289 $   & $ 2.68$ & $ 22.12$  \\ 
 \hline
 
  $  3.5$ & $  5.5$ & $ 216 $ & $2119 $ & $ 2093$  & $2.58 $ & $23.00 $  \\ 
 \hline

 \hline
\end{tabular}}
\caption{Statistics of superclusters found with `adaptive' threshold method: number of overdense regions (\nr) and \scls (\nc), filling factor ($ FF $), and fraction of galaxies in \scls ($f_{\rm gal}$) for different combinations of density threshold and volume limits $\lbrace \dthmin, \dthmax ~{\rm and}~ \vth \rbrace$.
Primary choice of the parameters is $\vth=1728 ~\mpc^3$, $\dthmin=3.5$, $\dthmax=5.5$ (indicated by boldface). 
}
\label{tab:adapt}
\end{table}

An alternative to fixing a density threshold for all superclusters, along with a need to justify the selection of a specific one, is to try to devise a rule to find a border condition for each \scl separately. 
This is akin to watershed methods used to determine large-scale voids \citep{Platen:2007,Nadathur:2014}. 
While not as detailed in drawing structure boundaries as those mentioned, we implement an adaptive scheme to derive individual thresholds for all superclusters. 
Closely following \cite{Juhan_thesis, Liivamagi:2010jg}, the adaptive threshold method is briefly described below. 

\begin{itemize}
    \item We scan the density field at multiple thresholds, spaced linearly by $\Delta D$ in the interval $\dthmin \leq \dth \leq \dthmax$.
    
    \item Like in the fixed threshold case, we again set a minimum volume for a supercluster $\vth$ which can be different from what was considered in the fixed threshold case.
    
    \item At each threshold, say $\dth=D_n$, we keep track of all the superclusters and also the threshold associated with each supercluster.
    Then we increase the threshold by $\Delta D$: $\dth \to D_{n+1}=D_n + \Delta D$. 
 
    \item Suppose we see that at $\dth=D_{n+1}$ a \scl splits into multiple isolated regions. If $N \geq 2$ of them have $V \geq \vth$, we allow the splitting and consider each of these N fragments as a separate supercluster and set their critical thresholds to $D_{n+1}$. 
    Otherwise, we do not allow splitting and keep considering the larger one as a single supercluster. 
 
    \item Similarly to the fixed threshold case, we impose the condition that each \scl should include at least two galaxies, discarding them otherwise.
\end{itemize}

\Scls found in this adaptive method depend on the choice of $\dthmin, \dthmax, \vth$ and $\Delta D$. 
However, we observe that the method is sensitive to the choice of $\dthmin$ and $\vth$, and very weakly dependent on $\dthmax$ (for sufficiently large $\dthmax$) and almost insensitive to $\Delta D$ (for sufficiently small $\Delta D$). 
The statistics of adaptive threshold \scls corresponding to different choices of $\dthmin$, $\dthmax$ and $\vth$ have been summarised in Table \ref{tab:adapt}, some interesting details are given in Appendix~\ref{app:adaptive}. 
Now we have to choose the suitable values of the parameters, most importantly $\dthmin$ and $\vth$. 
\citet{Juhan_thesis, Liivamagi:2010jg} consider a minimum `diameter' of twice the smoothing scale. 
However, for a more streamlined computational pipeline it will be more suitable to use a volume cutoff like in the fixed threshold case. 
The diameter threshold of \citet{Juhan_thesis, Liivamagi:2010jg} can be viewed as an equivalent of a volume cutoff at $\vth= a^3=1728 ~\mpc^3$ since the smoothing scale used in this work is $a=12~\mpc$.
 
\begin{figure}
\centering
\includegraphics[width=\linewidth]{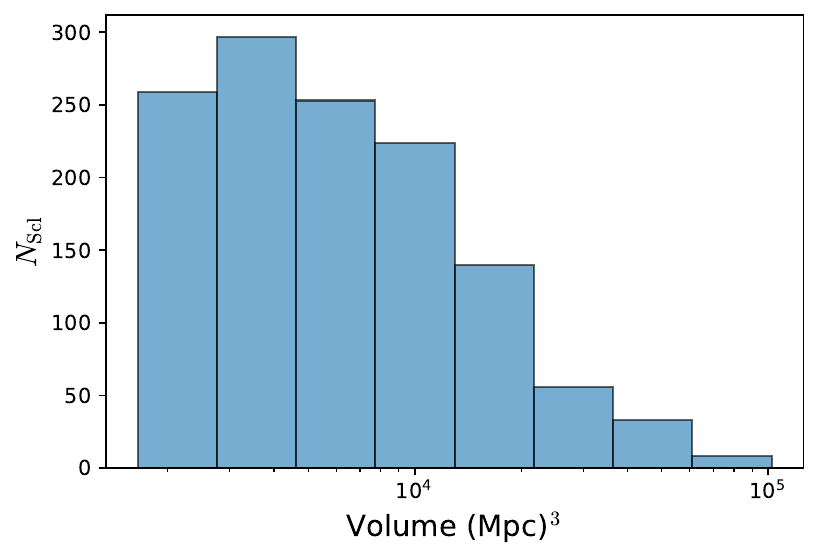}
\caption{Volume distribution of the adaptive threshold superclusters.}
\label{fig:Adaptive_vol_dist}
\end{figure}

Alongside, we set $\dthmin =3.5$, slightly above the percolation transition which takes place at $\dth \sim 2.5$.
To summarise, we set $\vth=1728 ~\mpc^3$ and $\dthmin=3.5$, $\dthmax=5.5$, $\Delta D=0.05$ in this work. 
This choice, which is consistent with that in \citet{Juhan_thesis, Liivamagi:2010jg}, has been highlighted in boldface in Table \ref{tab:adapt}. 
For this choice, the number of superclusters, \nc$=1271$, is slightly larger than that in our fixed threshold case, but we have now quite large \ff and $f_{\rm gal}$. The volume distribution of this adaptive threshold \scl sample has been shown in Figure~\ref{fig:Adaptive_vol_dist}. Note that we set the volume cutoff higher than that in the fixed threshold case. Still we find that most adaptive threshold \scls have volume a few thousand $\mpc^3$ which is consistent with that of the fixed threshold \scl sample.  

\subsection{Results with adaptive supercluster thresholds}

\begin{figure}
\centering
\includegraphics[width=\linewidth]{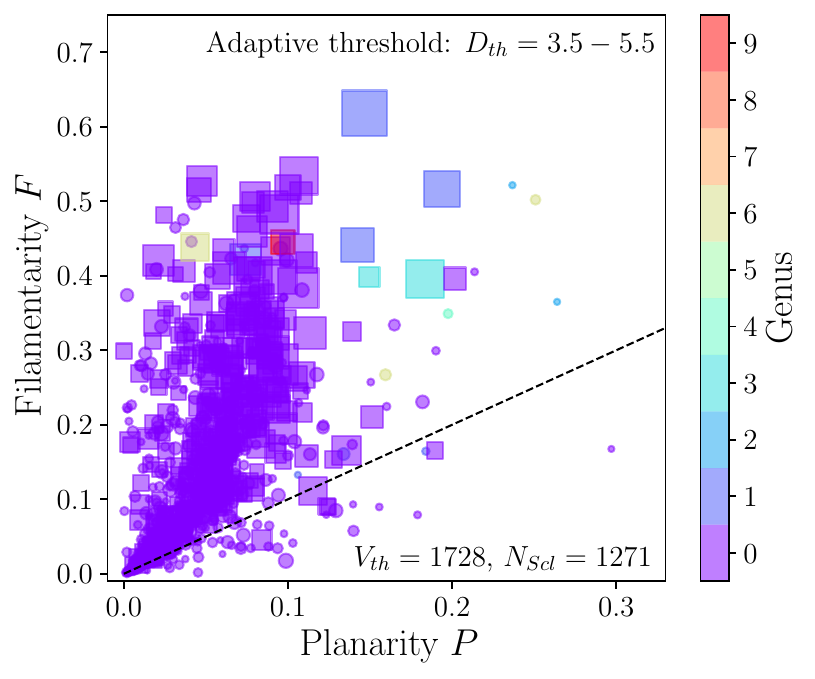}
\caption{Distribution of Shapefinders for adaptive threshold superclusters.
The smaller ($V < 10^{4}$ Mpc$^3$) and larger ($V \geq 10^{4}$ Mpc$^3$) \scls are marked by the filled circles and squares respectively where the size of the markers is proportional to the volume of the superclusters. The genus value of the \scls is indicated by the colour bar. The dashed line represents $F=P$ straight line above which most \scls lie similar to the fixed threshold case.}
\label{fig:Adaptive_SF_Dth_main}
\end{figure}

The planarity ($P$) and filamentarity ($F$) of the individual \scls defined in the adaptive threshold approach have been shown in Figure~\ref{fig:Adaptive_SF_Dth_main} for the parameters described above, namely $\vth=1728 ~\mpc^3$ and $\dthmin=3.5$, $\dthmax=5.5$, $\Delta D=0.05$. Similar to the fixed threshold case, filled circles and squares represent smaller and larger \scls where the size of the markers is proportional to the supercluster-volumes. 
Comparing with Figure~\ref{fig:Flat_SF_Dth5} we notice that the $P-F$ distribution of \scls for the adaptive threshold method is quite similar to that of the fixed threshold case. 
We again find that most smaller \scls are somewhat spherical ($P \approx 0 \approx F$) with zero genus while the large \scls can be quite filamentary. 
Furthermore, the outliers tend to have a non-trivial topology.

\begin{figure*}
\centering
\subfigure[]{
\includegraphics[width=0.465\textwidth]{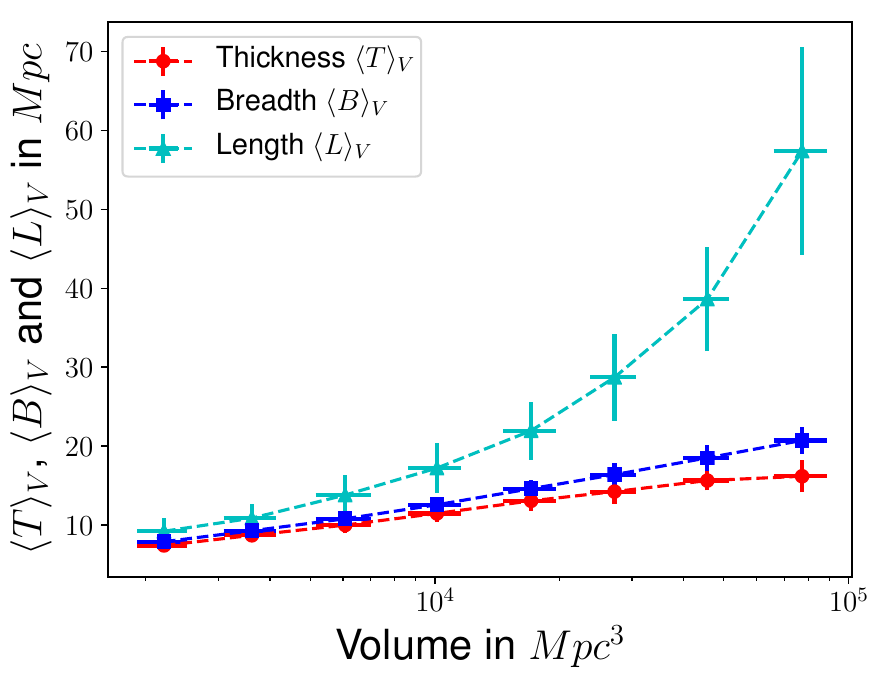}}
\subfigure[]{
\includegraphics[width=0.508\textwidth]{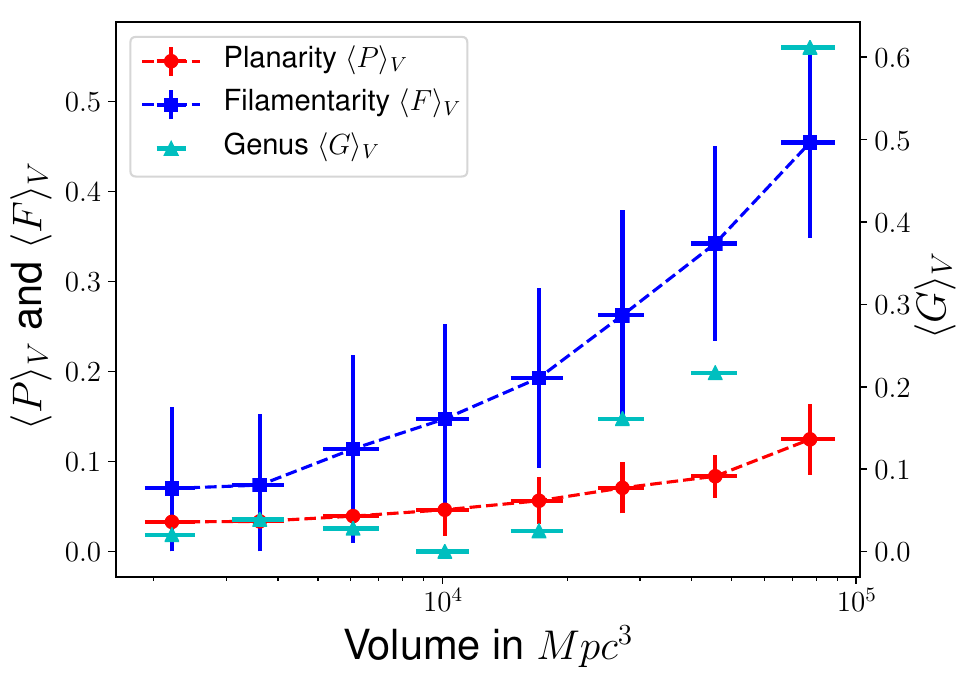}}
\caption{Volume-averaged Shapefinders of the superclusters with adaptive thresholds (extracted using $\lbrace \dthmin=3.5, ~\dthmax=5.5, ~\vth=1728 \rbrace$) are plotted against the volume of the \scls (divided into $8$ bins). Genus (volume-averaged in each bin) has been plotted along the right y-axis of the right panel. The error bars represent the scatter of the respective quantities (except for genus) in each bin.}
\label{fig:adaptive_SF_Dths_v}
\end{figure*}

 We show the dependency of the Shapefinders on the volume of \scls in Figure~\ref{fig:adaptive_SF_Dths_v}. 
 From the left panel, it is evident that the third Shapefinder (length) of the larger \scls is much larger than their first two Shapefinders (breadth and thickness), $L \gg B\sim T$, similar to what we observe for the fixed threshold supercluster sample. 
 This results in higher filamentarity among the large \scls while the planarity remains small and does not increase much as we move to higher volume bins, thus $F > P$ as shown by the right panel. 
 Again, the (volume-averaged) genus increases with the volume and $\langle G \rangle_V$ for large adaptive threshold \scls is significantly higher than that for the fixed threshold case (compare the right panel with that of Figure~\ref{fig:Flat_SF_Dth5_v}).

\begin{figure}
\centering
\includegraphics[width=\linewidth]{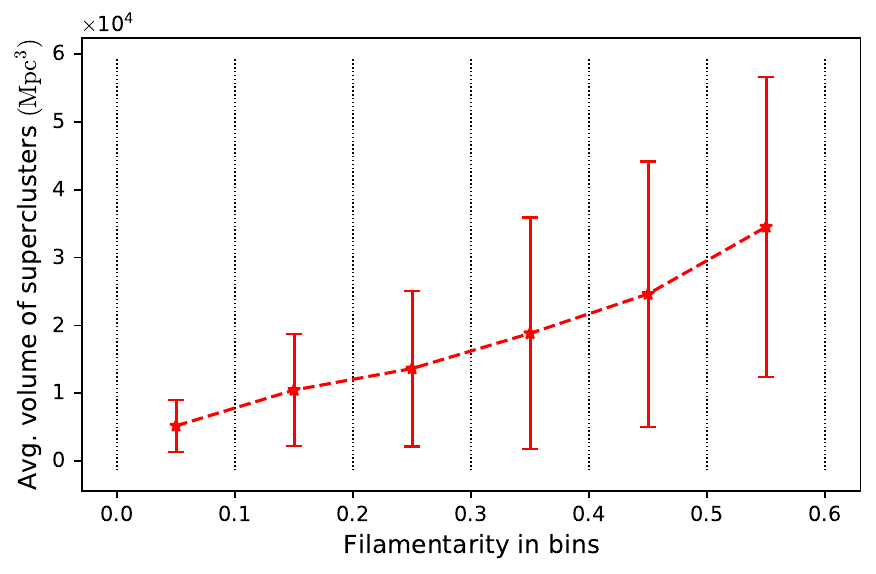}
\caption{Average volume the \scls belonging to different filamentarity bins has been shown for the adaptive threshold \scl sample (same as Figure~\ref{fig:Flat_F_bins} but for the adaptive case). The error bars represent the standard deviation in each bin.}
\label{fig:Adaptive_F_bins}
\end{figure}

Figure~\ref{fig:Adaptive_F_bins} shows the average volume of adaptive threshold \scls falling into different filamentarity bins. Again, we clearly notice that higher filamentary \scls (fewer in number) tend to be larger in size, similar to what we observe for the fixed threshold \scl samples.

\begin{figure*}
\centering
\subfigure[]{
\includegraphics[width=0.485\textwidth]{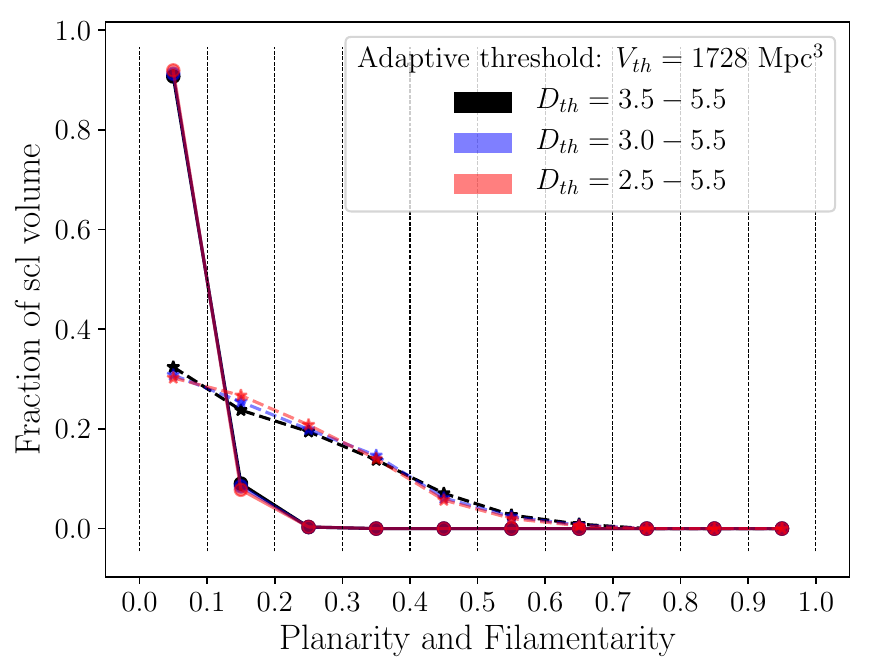}}
\subfigure[]{
\includegraphics[width=0.485\textwidth]{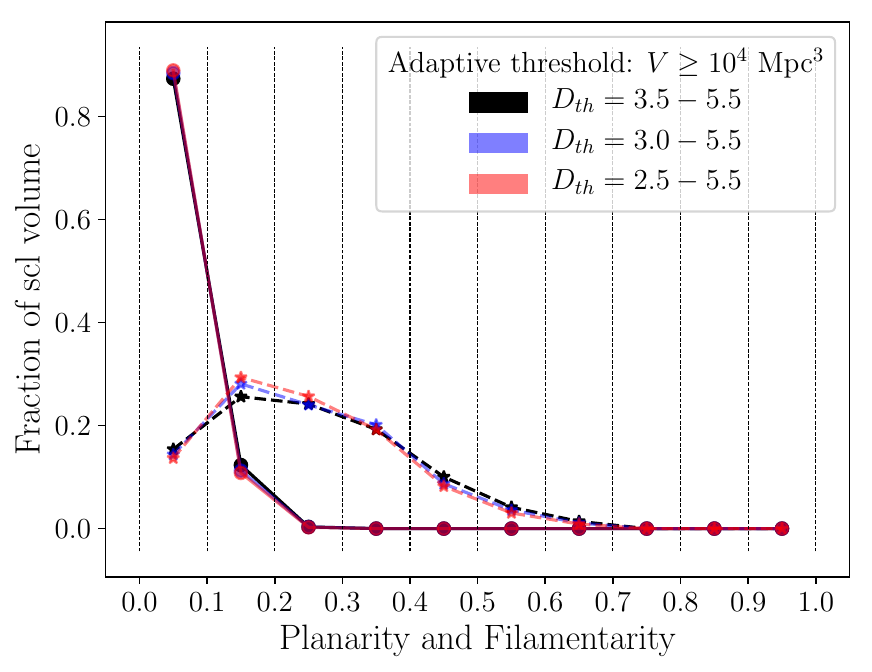}}
\caption{Volume weighted distributions of planarity (circles joined by solid lines) and filamentarity (stars joined by dashed lines) of  all adaptive threshold \scls is shown in the left. The right panel shows the same but only considering the large \scls with $V \geq 10^4 \mpc$. 
Different colours represent various choices of $\dthmin$.}
\label{fig:Adaptive_frac_SF}
\end{figure*}

The volume weighted shape distribution of adaptive threshold \scls is shown in Figure~\ref{fig:Adaptive_frac_SF}, for three different values of $\dthmin$ (the black curves represent our primary choice, $\dthmin=3.5$). 
The left panel includes all the \scls while only the large \scls are considered in the right panel. 
Along the y-axes of both panels, we plot the fraction of total supercluster volume falling inside different planarity/filamentarity bins, as explained in Eq.~\eqref{eq:wV_shape}. 
Remarkably, although different $\dthmin$ results in different numbers of \scls (and \ff) but the shape distributions for them are almost identical. From the right panel it is evident that \scls making up most of the volume have nonzero filamentarity dominated by the range $0.1 <F<0.2$. 
More figures comparing the shape distribution of adaptive threshold \scls are provided in Appendix~\ref{app:shape_dist_adap}. 

Finally, we compare the shape distribution of \scls obtained following the fixed and adaptive threshold approaches in Appendix~\ref{app:comp}. Although these two distinct definitions give rise to different \scl samples, we find that the shape distribution is very much alike. 

\section{Conclusions and discussion}
    
    Superclusters are dense and distinctive extended objects in the Universe and hence include crucial imprints of the structure formation process. 
    Nevertheless, lacking a unique definition they are often defined according to the problem in hand. 
    Here we study the superclusters which are extracted from a luminosity density field obtained from SDSS DR12 main galaxy sample. We construct \scls samples separately following two definitions, namely, (i) with a fixed density threshold and (ii) with adaptive, individually assigned thresholds. 
    After delineating the superclusters, we analyse the geometry of the individuals using the Minkowski functionals and their ratios, called Shapefinders.
    While \citet{Einasto:2011zc}, \citet{Einasto:2011} have studied the morphology of individual superclusters or smaller subsamples of superclusters in the SDSS main galaxy survey (usually within a limited redshift range), current work distinguishes itself by covering the full galaxy sample volume and more precise calculations of the Shapefinders. The important results we find from studying the morphology and shape distribution of the \scls are highlighted below. 

\begin{itemize}
    \item Although the number and filling factor of \scls found in these different approaches differ significantly, the morphologies are quite similar. 
    The shape distribution we present in this article depends very little on the convention of defining the \scls and therefore is quite robust.
    
    \item We prepare catalogues of shapes and morphology of superclusters defined in both fixed threshold and adaptive threshold methods. 
    The catalogues are released publicly with this work (see `Data Availability' section).  
    
    \item We found that smaller superclusters mostly have spherical shape ($P \approx 0 \approx F$) which should be influenced by the choice of the spherical smoothing filter while constructing the density field.

    \item However, the larger \scls ($V\gtrsim 10^4~\mpc^3 $) can be quite filamentary as well as multiply connected. Approximately $80\%$ of these large \scls (in both fixed and adaptive threshold samples) have $F>0.1$ (and roughly $40\%$ have $F>0.3$).  
    
    \item The \scls with spherical morphology make up most of the supercluster-volume. However, if we consider only large superclusters, the total supercluster-volume is dominated by mildly filamentary structures. For example, $\sim 88\%$  of the total volume of these large superclusters belongs to superclusters with $F>0.1$. Interestingly, we find that larger volume fraction is moderately filamentary in the adaptive threshold case as compared to the fixed threshold sample where slightly bigger volume fraction falls inside the higher filamentarity bins (see Figure~\ref{fig:Comp_frac_SF_cth10000}).

    \item The shape/morphology distributions found in this work (Figures~\ref{fig:Flat_SF_Dth5}, \ref{fig:Adaptive_SF_Dth_main}) are consistent with the earlier works on SDSS DR7 \cite{Liivamagi:2010jg}.
\end{itemize}

There are a lot of interesting aspects that can be pursued in the future. 
This work would allow us to shed light on the connections between the geometrical aspects of the large-scale structure and various properties related to structure formation. E.g. using the size and shape information of each \scl one can look into the correlations between supercluster shapes/morphologies and member galaxy properties (say star formation rate, colour etc); some earlier studies attempted similar exercise on a smaller supercluster sample \citep{M2014,Cohen2017}. Side by side, one can focus on the outlier \scls in $P-F$ diagrams and investigate how they are different from the rest.
Another important future aspect of the work can be comparisons of observed structures with the ones from various large-scales simulations (based on $\Lambda$CDM as well as different modified gravity models) from the geometrical points of view, that can be performed using the shape distribution of the superclusters. 
Though, it is essential to have a consistent definition and method of extraction for superclusters for this type of exercise. Note that, similar analysis can be performed on the voids too, however, delineating voids is a less straightforward exercise.

\section*{Acknowledgement}
SB thanks Elmo Tempel for the hospitality and support during a visit to
the Tartu Observatory, University of Tartu where the project was conceived and the early part of this work was done. 
SB also thanks Varun Sahni, Prakash Sarkar, Santanu Das for their contributions in developing SURFGEN2 in its initial phase.
We acknowledge the support by ETAg grant PRG1006 and by EU through the ERDF CoE grant TK133. We are grateful to the anonymous referee for the comments and suggestions which greatly improved the quality of this paper.

\section*{Data availability}

The SDSS DR12 data and the density field used in this article will be shared on reasonable request to the corresponding author. The supercluster shape and morphology catalogues can be accessed at \href{https://github.com/deltasata/Morphology_SDSS_DR12_scls}{https://github.com/deltasata/Morphology\_SDSS\_DR12\_scls}.

\bibliographystyle{mnras}
\bibliography{sdss_morphology}

\begin{thebibliography}{}
\makeatletter
\relax
\def\mn@urlcharsother{\let\do\@makeother \do\$\do\&\do\#\do\^\do\_\do\%\do\~}
\def\mn@doi{\begingroup\mn@urlcharsother \@ifnextchar [ {\mn@doi@}
  {\mn@doi@[]}}
\def\mn@doi@[#1]#2{\def\@tempa{#1}\ifx\@tempa\@empty \href
  {http://dx.doi.org/#2} {doi:#2}\else \href {http://dx.doi.org/#2} {#1}\fi
  \endgroup}
\def\mn@eprint#1#2{\mn@eprint@#1:#2::\@nil}
\def\mn@eprint@arXiv#1{\href {http://arxiv.org/abs/#1} {{\tt arXiv:#1}}}
\def\mn@eprint@dblp#1{\href {http://dblp.uni-trier.de/rec/bibtex/#1.xml}
  {dblp:#1}}
\def\mn@eprint@#1:#2:#3:#4\@nil{\def\@tempa {#1}\def\@tempb {#2}\def\@tempc
  {#3}\ifx \@tempc \@empty \let \@tempc \@tempb \let \@tempb \@tempa \fi \ifx
  \@tempb \@empty \def\@tempb {arXiv}\fi \@ifundefined
  {mn@eprint@\@tempb}{\@tempb:\@tempc}{\expandafter \expandafter \csname
  mn@eprint@\@tempb\endcsname \expandafter{\@tempc}}}

\bibitem[\protect\citeauthoryear{{Ahumada} et~al.,}{{Ahumada}
  et~al.}{2020}]{Ahumada:2020dr16}
{Ahumada} R.,  et~al., 2020, \mn@doi [\apjs] {10.3847/1538-4365/ab929e}, \href
  {https://ui.adsabs.harvard.edu/abs/2020ApJS..249....3A} {249, 3}

\bibitem[\protect\citeauthoryear{{Alam} et~al.,}{{Alam}
  et~al.}{2015}]{Alam:2015dr12}
{Alam} S.,  et~al., 2015, \mn@doi [\apjs] {10.1088/0067-0049/219/1/12}, \href
  {https://ui.adsabs.harvard.edu/abs/2015ApJS..219...12A} {219, 12}

\bibitem[\protect\citeauthoryear{Appleby, Park, Pranav, Hong, Hwang, Kim  \&
  Buchert}{Appleby et~al.}{2021}]{Appleby:2021xoz}
Appleby S.,  Park C.,  Pranav P.,  Hong S.~E.,  Hwang H.~S.,  Kim J.,   Buchert
  T.,  2021, arXiv:2110.06109 [astro-ph.CO]

\bibitem[\protect\citeauthoryear{Bag, Mondal, Sarkar, Bharadwaj  \& Sahni}{Bag
  et~al.}{2018}]{Bag:2018zon}
Bag S.,  Mondal R.,  Sarkar P.,  Bharadwaj S.,   Sahni V.,  2018, \mn@doi [Mon.
  Not. Roy. Astron. Soc.] {10.1093/mnras/sty714}, 477, 1984

\bibitem[\protect\citeauthoryear{Bag, Mondal, Sarkar, Bharadwaj, Choudhury  \&
  Sahni}{Bag et~al.}{2019}]{Bag:2018fyr}
Bag S.,  Mondal R.,  Sarkar P.,  Bharadwaj S.,  Choudhury T.~R.,   Sahni V.,
  2019, \mn@doi [Mon. Not. Roy. Astron. Soc.] {10.1093/mnras/stz532}, 485, 2235

\bibitem[\protect\citeauthoryear{{Bagchi}, {Sankhyayan}, {Sarkar},
  {Raychaudhury}, {Jacob}  \& {Dabhade}}{{Bagchi}
  et~al.}{2017}]{2017ApJ...844...25B}
{Bagchi} J.,  {Sankhyayan} S.,  {Sarkar} P.,  {Raychaudhury} S.,  {Jacob} J.,
  {Dabhade} P.,  2017, \mn@doi [\apj] {10.3847/1538-4357/aa7949}, \href
  {https://ui.adsabs.harvard.edu/abs/2017ApJ...844...25B} {844, 25}

\bibitem[\protect\citeauthoryear{Bernardeau, Colombi, Gaztanaga  \&
  Scoccimarro}{Bernardeau et~al.}{2002}]{Bernardeau:2001qr}
Bernardeau F.,  Colombi S.,  Gaztanaga E.,   Scoccimarro R.,  2002, \mn@doi
  [Phys. Rept.] {10.1016/S0370-1573(02)00135-7}, 367, 1

\bibitem[\protect\citeauthoryear{{Bond}, {Kofman}  \& {Pogosyan}}{{Bond}
  et~al.}{1996}]{1996Natur.380..603B}
{Bond} J.~R.,  {Kofman} L.,   {Pogosyan} D.,  1996, \mn@doi [\nat]
  {10.1038/380603a0}, \href
  {https://ui.adsabs.harvard.edu/abs/1996Natur.380..603B} {380, 603}

\bibitem[\protect\citeauthoryear{{Bond}, {Strauss}  \& {Cen}}{{Bond}
  et~al.}{2010}]{2010MNRAS.409..156B}
{Bond} N.~A.,  {Strauss} M.~A.,   {Cen} R.,  2010, \mn@doi [\mnras]
  {10.1111/j.1365-2966.2010.17307.x}, \href
  {https://ui.adsabs.harvard.edu/abs/2010MNRAS.409..156B} {409, 156}

\bibitem[\protect\citeauthoryear{Chen, Xu, Wang  \& Chen}{Chen
  et~al.}{2019}]{Chen:2018enj}
Chen Z.,  Xu Y.,  Wang Y.,   Chen X.,  2019, \mn@doi [Astrophys. J.]
  {10.3847/1538-4357/ab43e6}, 885, 23

\bibitem[\protect\citeauthoryear{{Chernyaev}}{{Chernyaev}}{1987}]{marcube}
{Chernyaev} E.~V.,  1987, \mn@doi [ACM SIGGRAPH Computer Graphics]
  {10.1145/37402.37422}, 21, 163

\bibitem[\protect\citeauthoryear{{Cohen}, {Hickox}, {Wegner}, {Einasto}  \&
  {Vennik}}{{Cohen} et~al.}{2017}]{Cohen2017}
{Cohen} S.~A.,  {Hickox} R.~C.,  {Wegner} G.~A.,  {Einasto} M.,   {Vennik} J.,
  2017, \mn@doi [\apj] {10.3847/1538-4357/835/1/56}, \href
  {https://ui.adsabs.harvard.edu/abs/2017ApJ...835...56C} {835, 56}

\bibitem[\protect\citeauthoryear{Crofton}{Crofton}{1868}]{Crofton_1868}
Crofton M.~W.,  1868, Phil. Trans. R. Soc. Lond, 158, 181

\bibitem[\protect\citeauthoryear{{Davis}, {Efstathiou}, {Frenk}  \&
  {White}}{{Davis} et~al.}{1985}]{1985ApJ...292..371D}
{Davis} M.,  {Efstathiou} G.,  {Frenk} C.~S.,   {White} S.~D.~M.,  1985,
  \mn@doi [\apj] {10.1086/163168}, \href
  {https://ui.adsabs.harvard.edu/abs/1985ApJ...292..371D} {292, 371}

\bibitem[\protect\citeauthoryear{{Davison} \& {Hinkley}}{{Davison} \&
  {Hinkley}}{1997}]{davison1997}
{Davison} A.~C.,  {Hinkley} D.~V.,  1997, {Bootstrap Methods and Their
  Application}.
Cambridge University Press, Cambridge, UK

\bibitem[\protect\citeauthoryear{{Einasto}, {Klypin}, {Saar}  \&
  {Shandarin}}{{Einasto} et~al.}{1984}]{1984MNRAS.206..529E}
{Einasto} J.,  {Klypin} A.~A.,  {Saar} E.,   {Shandarin} S.~F.,  1984, \mn@doi
  [\mnras] {10.1093/mnras/206.3.529}, \href
  {https://ui.adsabs.harvard.edu/abs/1984MNRAS.206..529E} {206, 529}

\bibitem[\protect\citeauthoryear{{Einasto} et~al.,}{{Einasto}
  et~al.}{2007}]{M2007A}
{Einasto} M.,  et~al., 2007, \mn@doi [\aap] {10.1051/0004-6361:20078037}, \href
  {https://ui.adsabs.harvard.edu/abs/2007A&A...476..697E} {476, 697}

\bibitem[\protect\citeauthoryear{Einasto, Liivamagi, Tago, Saar, Tempel,
  Einasto, Martinez  \& Heinamaki}{Einasto et~al.}{2011a}]{Einasto:2011zc}
Einasto M.,  Liivamagi L.,  Tago E.,  Saar E.,  Tempel E.,  Einasto J.,
  Martinez V.,   Heinamaki P.,  2011a, \mn@doi [Astron. Astrophys.]
  {10.1051/0004-6361/201116564}, 532, A5

\bibitem[\protect\citeauthoryear{{Einasto}, {Liivam{\"a}gi}, {Saar}, {Einasto},
  {Tempel}, {Tago}  \& {Mart{\'\i}nez}}{{Einasto} et~al.}{2011b}]{M2011c}
{Einasto} M.,  {Liivam{\"a}gi} L.~J.,  {Saar} E.,  {Einasto} J.,  {Tempel} E.,
  {Tago} E.,   {Mart{\'\i}nez} V.~J.,  2011b, \mn@doi [\aap]
  {10.1051/0004-6361/201117529}, \href
  {https://ui.adsabs.harvard.edu/abs/2011A&A...535A..36E} {535, A36}

\bibitem[\protect\citeauthoryear{{Einasto} et~al.,}{{Einasto}
  et~al.}{2011c}]{Einasto:2011}
{Einasto} M.,  et~al., 2011c, \mn@doi [\apj] {10.1088/0004-637X/736/1/51},
  \href {http://adsabs.harvard.edu/abs/2011ApJ...736...51E} {736, 51}

\bibitem[\protect\citeauthoryear{{Einasto}, {Lietzen}, {Tempel}, {Gramann},
  {Liivam{\"a}gi}  \& {Einasto}}{{Einasto} et~al.}{2014}]{M2014}
{Einasto} M.,  {Lietzen} H.,  {Tempel} E.,  {Gramann} M.,  {Liivam{\"a}gi}
  L.~J.,   {Einasto} J.,  2014, \mn@doi [\aap] {10.1051/0004-6361/201323111},
  \href {https://ui.adsabs.harvard.edu/abs/2014A&A...562A..87E} {562, A87}

\bibitem[\protect\citeauthoryear{{Einasto} et~al.,}{{Einasto}
  et~al.}{2015}]{2015A&A...580A..69E}
{Einasto} M.,  et~al., 2015, \mn@doi [\aap] {10.1051/0004-6361/201526399},
  \href {http://adsabs.harvard.edu/abs/2015A%26A...580A..69E} {580, A69}

\bibitem[\protect\citeauthoryear{{Einasto} et~al.,}{{Einasto}
  et~al.}{2016}]{2016A&A...595A..70E}
{Einasto} M.,  et~al., 2016, \mn@doi [\aap] {10.1051/0004-6361/201628567},
  \href {http://adsabs.harvard.edu/abs/2016A%26A...595A..70E} {595, A70}

\bibitem[\protect\citeauthoryear{{Einasto}, {Suhhonenko}, {Liivam{\"a}gi}  \&
  {Einasto}}{{Einasto} et~al.}{2018}]{J2018}
{Einasto} J.,  {Suhhonenko} I.,  {Liivam{\"a}gi} L.~J.,   {Einasto} M.,  2018,
  \mn@doi [\aap] {10.1051/0004-6361/201833011}, \href
  {https://ui.adsabs.harvard.edu/abs/2018A&A...616A.141E} {616, A141}

\bibitem[\protect\citeauthoryear{{Einasto}, {Suhhonenko}, {Liivam{\"a}gi}  \&
  {Einasto}}{{Einasto} et~al.}{2019}]{J2019}
{Einasto} J.,  {Suhhonenko} I.,  {Liivam{\"a}gi} L.~J.,   {Einasto} M.,  2019,
  \mn@doi [\aap] {10.1051/0004-6361/201834450}, \href
  {https://ui.adsabs.harvard.edu/abs/2019A&A...623A..97E} {623, A97}

\bibitem[\protect\citeauthoryear{{Einasto} et~al.,}{{Einasto}
  et~al.}{2020}]{Einasto2020}
{Einasto} M.,  et~al., 2020, \mn@doi [\aap] {10.1051/0004-6361/202037982},
  \href {https://ui.adsabs.harvard.edu/abs/2020A&A...641A.172E} {641, A172}

\bibitem[\protect\citeauthoryear{{Einasto}, {H{\"u}tsi}, {Suhhonenko},
  {Liivam{\"a}gi}  \& {Einasto}}{{Einasto} et~al.}{2021a}]{J2021}
{Einasto} J.,  {H{\"u}tsi} G.,  {Suhhonenko} I.,  {Liivam{\"a}gi} L.~J.,
  {Einasto} M.,  2021a, \mn@doi [\aap] {10.1051/0004-6361/202038358}, \href
  {https://ui.adsabs.harvard.edu/abs/2021A&A...647A..17E} {647, A17}

\bibitem[\protect\citeauthoryear{{Einasto} et~al.,}{{Einasto}
  et~al.}{2021b}]{2021A&A...649A..51E}
{Einasto} M.,  et~al., 2021b, \mn@doi [\aap] {10.1051/0004-6361/202040200},
  \href {https://ui.adsabs.harvard.edu/abs/2021A&A...649A..51E} {649, A51}

\bibitem[\protect\citeauthoryear{{Einasto} et~al.,}{{Einasto}
  et~al.}{2022}]{2022arXiv220408918E}
{Einasto} M.,  et~al., 2022, arXiv e-prints, \href
  {https://ui.adsabs.harvard.edu/abs/2022arXiv220408918E} {p. arXiv:2204.08918}

\bibitem[\protect\citeauthoryear{{Eisenstein} et~al.,}{{Eisenstein}
  et~al.}{2011}]{Eisenstein:2011sdss}
{Eisenstein} D.~J.,  et~al., 2011, \mn@doi [\aj] {10.1088/0004-6256/142/3/72},
  \href {https://ui.adsabs.harvard.edu/abs/2011AJ....142...72E} {142, 72}

\bibitem[\protect\citeauthoryear{{Friedrich}, {Mellema}, {Alvarez}, {Shapiro}
  \& {Iliev}}{{Friedrich} et~al.}{2011}]{Friedrich2011}
{Friedrich} M.~M.,  {Mellema} G.,  {Alvarez} M.~A.,  {Shapiro} P.~R.,   {Iliev}
  I.~T.,  2011, \mn@doi [\mnras] {10.1111/j.1365-2966.2011.18219.x}, \href
  {http://adsabs.harvard.edu/abs/2011MNRAS.413.1353F} {413, 1353}

\bibitem[\protect\citeauthoryear{Gorbunov \& Rubakov}{Gorbunov \&
  Rubakov}{2011}]{2011iteu.book.....G}
Gorbunov D.~S.,  Rubakov V.~A.,  2011, {Introduction to the theory of the early
  universe: cosmological perturbations and inflationary theory}.
World Scientific, \url {https://cds.cern.ch/record/1354521}

\bibitem[\protect\citeauthoryear{{Gregory} \& {Thompson}}{{Gregory} \&
  {Thompson}}{1978}]{1978ApJ...222..784G}
{Gregory} S.~A.,  {Thompson} L.~A.,  1978, \mn@doi [\apj] {10.1086/156198},
  \href {https://ui.adsabs.harvard.edu/abs/1978ApJ...222..784G} {222, 784}

\bibitem[\protect\citeauthoryear{{Hikage} et~al.,}{{Hikage}
  et~al.}{2003}]{Hikage2003}
{Hikage} C.,  et~al., 2003, \mn@doi [\pasj] {10.1093/pasj/55.5.911}, \href
  {http://adsabs.harvard.edu/abs/2003PASJ...55..911H} {55, 911}

\bibitem[\protect\citeauthoryear{{Hikage}, {Komatsu}  \& {Matsubara}}{{Hikage}
  et~al.}{2006}]{Hikage2006}
{Hikage} C.,  {Komatsu} E.,   {Matsubara} T.,  2006, \mn@doi [\apj]
  {10.1086/508653}, \href {http://adsabs.harvard.edu/abs/2006ApJ...653...11H}
  {653, 11}

\bibitem[\protect\citeauthoryear{{J{\~o}eveer}, {Einasto}  \&
  {Tago}}{{J{\~o}eveer} et~al.}{1978}]{1978MNRAS.185..357J}
{J{\~o}eveer} M.,  {Einasto} J.,   {Tago} E.,  1978, \mn@doi [\mnras]
  {10.1093/mnras/185.2.357}, \href
  {https://ui.adsabs.harvard.edu/abs/1978MNRAS.185..357J} {185, 357}

\bibitem[\protect\citeauthoryear{Jasche, Leclercq  \& Wandelt}{Jasche
  et~al.}{2015}]{Jasche:2014vpa}
Jasche J.,  Leclercq F.,   Wandelt B.~D.,  2015, \mn@doi [JCAP]
  {10.1088/1475-7516/2015/01/036}, 01, 036

\bibitem[\protect\citeauthoryear{Kaiser}{Kaiser}{1987}]{Kaiser:1987qv}
Kaiser N.,  1987, Mon. Not. Roy. Astron. Soc., 227, 1

\bibitem[\protect\citeauthoryear{Kapahtia, Chingangbam, Ghara, Appleby  \&
  Choudhury}{Kapahtia et~al.}{2021}]{Kapahtia:2021eok}
Kapahtia A.,  Chingangbam P.,  Ghara R.,  Appleby S.,   Choudhury T.~R.,  2021,
  arXiv:2101.03962 [astro-ph.CO]

\bibitem[\protect\citeauthoryear{Koenderink}{Koenderink}{1984}]{Koenderink}
Koenderink J.~J.,  1984, \mn@doi [Biological Cybernetics] {10.1007/BF00336961},
  50, 363

\bibitem[\protect\citeauthoryear{Kolb \& Turner}{Kolb \&
  Turner}{1990}]{1990eaun.book.....K}
Kolb E.~W.,  Turner M.~S.,  1990, {The early universe}.
Frontiers in physics, Westview Press, \mn@doi{10.1201/9780429492860}, \url
  {https://cds.cern.ch/record/206230}

\bibitem[\protect\citeauthoryear{{Libeskind} et~al.,}{{Libeskind}
  et~al.}{2018}]{Libeskind18}
{Libeskind} N.~I.,  et~al., 2018, \mn@doi [\mnras] {10.1093/mnras/stx1976},
  \href {https://ui.adsabs.harvard.edu/abs/2018MNRAS.473.1195L} {473, 1195}

\bibitem[\protect\citeauthoryear{{Lietzen} et~al.,}{{Lietzen}
  et~al.}{2016}]{2016A&A...588L...4L}
{Lietzen} H.,  et~al., 2016, \mn@doi [\aap] {10.1051/0004-6361/201628261},
  \href {http://adsabs.harvard.edu/abs/2016A%26A...588L...4L} {588, L4}

\bibitem[\protect\citeauthoryear{{Liivam{\"a}gi}}{{Liivam{\"a}gi}}{2017}]{Juhan_thesis}
{Liivam{\"a}gi} L.~J.,  2017, PhD thesis, Universitatis Tartuensis

\bibitem[\protect\citeauthoryear{Liivamagi, Tempel  \& Saar}{Liivamagi
  et~al.}{2012}]{Liivamagi:2010jg}
Liivamagi L.~J.,  Tempel E.,   Saar E.,  2012, \mn@doi [Astron. Astrophys.]
  {10.1051/0004-6361/201016288}, 539, A80

\bibitem[\protect\citeauthoryear{Lippich \& S'anchez}{Lippich \&
  S'anchez}{2020}]{Lippich:2020vpy}
Lippich M.,  S'anchez A.~G.,  2020, arXiv:2012.08529 [astro-ph.CO]

\bibitem[\protect\citeauthoryear{Lorensen \& Cline}{Lorensen \&
  Cline}{1995}]{mar33}
Lorensen W.~E.,  Cline H.~E.,  1995, Technical Report CERN-CN-95-17

\bibitem[\protect\citeauthoryear{Matsubara \& Kuriki}{Matsubara \&
  Kuriki}{2020}]{Matsubara:2020fet}
Matsubara T.,  Kuriki S.,  2020, arXiv:2011.04954 [astro-ph.CO]

\bibitem[\protect\citeauthoryear{Matsubara, Hikage  \& Kuriki}{Matsubara
  et~al.}{2020}]{Matsubara:2020knr}
Matsubara T.,  Hikage C.,   Kuriki S.,  2020, arXiv:2012.00203 [astro-ph.CO]

\bibitem[\protect\citeauthoryear{{Mecke}, {Buchert}  \& {Wagner}}{{Mecke}
  et~al.}{1994}]{mecke}
{Mecke} K.~R.,  {Buchert} T.,   {Wagner} H.,  1994, \aap, \href
  {http://adsabs.harvard.edu/abs/1994A%26A...288..697M} {288, 697}

\bibitem[\protect\citeauthoryear{Melott}{Melott}{1990}]{MELOTT19901}
Melott A.~L.,  1990, \mn@doi [Physics Reports]
  {https://doi.org/10.1016/0370-1573(90)90162-U}, 193, 1

\bibitem[\protect\citeauthoryear{{Nadathur} \& {Hotchkiss}}{{Nadathur} \&
  {Hotchkiss}}{2014}]{Nadathur:2014}
{Nadathur} S.,  {Hotchkiss} S.,  2014, \mn@doi [\mnras] {10.1093/mnras/stu349},
  \href {https://ui.adsabs.harvard.edu/abs/2014MNRAS.440.1248N} {440, 1248}

\bibitem[\protect\citeauthoryear{Nakahara}{Nakahara}{2003}]{Nakahara:2003nw}
Nakahara M.,  2003, Geometry, topology and physics.
\url {http://www.slac.stanford.edu/spires/find/hep/www?key=7208855}

\bibitem[\protect\citeauthoryear{{Novikov}, {Feldman}  \&
  {Shandarin}}{{Novikov} et~al.}{1999}]{Novikov1999}
{Novikov} D.,  {Feldman} H.~A.,   {Shandarin} S.~F.,  1999, \mn@doi
  [International Journal of Modern Physics D] {10.1142/S0218271899000225},
  \href {http://adsabs.harvard.edu/abs/1999IJMPD...8..291N} {8, 291}

\bibitem[\protect\citeauthoryear{{Novikov}, {Schmalzing}  \&
  {Mukhanov}}{{Novikov} et~al.}{2000}]{Novikov2000}
{Novikov} D.,  {Schmalzing} J.,   {Mukhanov} V.~F.,  2000, \aap, \href
  {http://adsabs.harvard.edu/abs/2000A%26A...364...17N} {364, 17}

\bibitem[\protect\citeauthoryear{{Pathak}, {Bag}, {Majumdar}, {Mondal},
  {Kamran}  \& {Sarkar}}{{Pathak} et~al.}{2022}]{Pathak:2022ahj}
{Pathak} A.,  {Bag} S.,  {Majumdar} S.,  {Mondal} R.,  {Kamran} M.,   {Sarkar}
  P.,  2022, arXiv e-prints, \href
  {https://ui.adsabs.harvard.edu/abs/2022arXiv220203701P} {p. arXiv:2202.03701}

\bibitem[\protect\citeauthoryear{{Planck Collaboration} et~al.,}{{Planck
  Collaboration} et~al.}{2016}]{Planck:2016VIII}
{Planck Collaboration} et~al., 2016, \mn@doi [\aap]
  {10.1051/0004-6361/201525830}, \href
  {https://ui.adsabs.harvard.edu/abs/2016A&A...594A..13P} {594, A13}

\bibitem[\protect\citeauthoryear{{Platen}, {van de Weygaert}  \&
  {Jones}}{{Platen} et~al.}{2007}]{Platen:2007}
{Platen} E.,  {van de Weygaert} R.,   {Jones} B. J.~T.,  2007, \mn@doi [\mnras]
  {10.1111/j.1365-2966.2007.12125.x}, \href
  {https://ui.adsabs.harvard.edu/abs/2007MNRAS.380..551P} {380, 551}

\bibitem[\protect\citeauthoryear{{Pratten} \& {Munshi}}{{Pratten} \&
  {Munshi}}{2012}]{Pratten}
{Pratten} G.,  {Munshi} D.,  2012, \mn@doi [\mnras]
  {10.1111/j.1365-2966.2012.21103.x}, \href
  {https://ui.adsabs.harvard.edu/abs/2012MNRAS.423.3209P} {423, 3209}

\bibitem[\protect\citeauthoryear{{Saar}}{{Saar}}{2009}]{saar2009}
{Saar} E.,  2009, in {V.~J.~Mart{\'{\i}}nez, E.~Saar,
  E.~Mart{\'{\i}}nez-Gonz{\'a}lez, \& M.-J.~Pons-Border{\'{\i}}a} ed.,  Lecture
  Notes in Physics, Berlin Springer Verlag Vol. 665, Data Analysis in
  Cosmology. pp 523--563, \mn@doi{10.1007/978-3-540-44767-2_16}

\bibitem[\protect\citeauthoryear{{Sahni} \& {Coles}}{{Sahni} \&
  {Coles}}{1995}]{1995PhR...262....1S}
{Sahni} V.,  {Coles} P.,  1995, \mn@doi [\physrep]
  {10.1016/0370-1573(95)00014-8}, \href
  {https://ui.adsabs.harvard.edu/abs/1995PhR...262....1S} {262, 1}

\bibitem[\protect\citeauthoryear{Sahni, Sathyaprakash  \& Shandarin}{Sahni
  et~al.}{1998}]{Sahni:1998cr}
Sahni V.,  Sathyaprakash B.,   Shandarin S.~F.,  1998, \mn@doi [Astrophys. J.
  Lett.] {10.1086/311214}, 495, L5

\bibitem[\protect\citeauthoryear{{Sathyaprakash}, {Sahni}  \&
  {Shandarin}}{{Sathyaprakash} et~al.}{1996}]{1996ApJ...462L...5S}
{Sathyaprakash} B.~S.,  {Sahni} V.,   {Shandarin} S.~F.,  1996, \mn@doi [\apjl]
  {10.1086/310024}, \href
  {https://ui.adsabs.harvard.edu/abs/1996ApJ...462L...5S} {462, L5}

\bibitem[\protect\citeauthoryear{Sathyaprakash, Sahni  \&
  Shandarin}{Sathyaprakash et~al.}{1998}]{Sathyaprakash:1998gy}
Sathyaprakash B.~S.,  Sahni V.,   Shandarin S.~F.,  1998, \mn@doi [Astrophys.
  J.] {10.1086/306447}, 508, 551

\bibitem[\protect\citeauthoryear{{Schmalzing} \& {Buchert}}{{Schmalzing} \&
  {Buchert}}{1997}]{Schmalzing1997}
{Schmalzing} J.,  {Buchert} T.,  1997, \mn@doi [\apjl] {10.1086/310680}, \href
  {http://adsabs.harvard.edu/abs/1997ApJ...482L...1S} {482, L1}

\bibitem[\protect\citeauthoryear{{Schmalzing} \& {Gorski}}{{Schmalzing} \&
  {Gorski}}{1998}]{Schmalzing1998}
{Schmalzing} J.,  {Gorski} K.~M.,  1998, \mn@doi [\mnras]
  {10.1046/j.1365-8711.1998.01467.x}, \href
  {http://adsabs.harvard.edu/abs/1998MNRAS.297..355S} {297, 355}

\bibitem[\protect\citeauthoryear{Schmalzing, Kerscher  \& Buchert}{Schmalzing
  et~al.}{1996}]{Schmalzing:1995qn}
Schmalzing J.,  Kerscher M.,   Buchert T.,  1996, \mn@doi [Proc. Int. Sch.
  Phys. Fermi] {10.3254/978-1-61499-217-2-281}, 132, 281

\bibitem[\protect\citeauthoryear{Shandarin \& Yess}{Shandarin \&
  Yess}{1998}]{Shandarin:1997fc}
Shandarin S.~F.,  Yess C.,  1998, \mn@doi [Astrophys. J.] {10.1086/306135},
  505, 12

\bibitem[\protect\citeauthoryear{Shandarin, Sheth  \& Sahni}{Shandarin
  et~al.}{2004}]{Shandarin:2003tx}
Shandarin S.~F.,  Sheth J.~V.,   Sahni V.,  2004, \mn@doi [Mon. Not. Roy.
  Astron. Soc.] {10.1111/j.1365-2966.2004.08060.x}, 353, 162

\bibitem[\protect\citeauthoryear{{Shapley}}{{Shapley}}{1930}]{1930BHarO.874....9S}
{Shapley} H.,  1930, Harvard College Observatory Bulletin, \href
  {https://ui.adsabs.harvard.edu/abs/1930BHarO.874....9S} {874, 9}

\bibitem[\protect\citeauthoryear{Sheth}{Sheth}{2004}]{Sheth:2003vm}
Sheth J.~V.,  2004, \mn@doi [Mon. Not. Roy. Astron. Soc.]
  {10.1111/j.1365-2966.2004.08191.x}, 354, 332

\bibitem[\protect\citeauthoryear{Sheth}{Sheth}{2006}]{Sheth:2006qz}
Sheth J.~V.,  2006, PhD thesis (\mn@eprint {arXiv} {astro-ph/0602433})

\bibitem[\protect\citeauthoryear{Sheth, Sahni, Shandarin  \&
  Sathyaprakash}{Sheth et~al.}{2003}]{Sheth:2002rf}
Sheth J.~V.,  Sahni V.,  Shandarin S.~F.,   Sathyaprakash B.,  2003, \mn@doi
  [Mon. Not. Roy. Astron. Soc.] {10.1046/j.1365-8711.2003.06642.x}, 343, 22

\bibitem[\protect\citeauthoryear{{Springel} et~al.,}{{Springel}
  et~al.}{2005}]{2005Natur.435..629S}
{Springel} V.,  et~al., 2005, \mn@doi [\nat] {10.1038/nature03597}, \href
  {https://ui.adsabs.harvard.edu/abs/2005Natur.435..629S} {435, 629}

\bibitem[\protect\citeauthoryear{Springel, Frenk  \& White}{Springel
  et~al.}{2006}]{Springel:2006vs}
Springel V.,  Frenk C.~S.,   White S. D.~M.,  2006, \mn@doi [Nature]
  {10.1038/nature04805}, 440, 1137

\bibitem[\protect\citeauthoryear{{Tempel} et~al.,}{{Tempel}
  et~al.}{2014}]{Tempel:2014gr}
{Tempel} E.,  et~al., 2014, \mn@doi [\aap] {10.1051/0004-6361/201423585}, \href
  {https://ui.adsabs.harvard.edu/abs/2014A&A...566A...1T} {566, A1}

\bibitem[\protect\citeauthoryear{{Tempel}, {Tuvikene}, {Kipper}  \&
  {Libeskind}}{{Tempel} et~al.}{2017}]{Tempel:2017gr}
{Tempel} E.,  {Tuvikene} T.,  {Kipper} R.,   {Libeskind} N.~I.,  2017, \mn@doi
  [\aap] {10.1051/0004-6361/201730499}, \href
  {https://ui.adsabs.harvard.edu/abs/2017A&A...602A.100T} {602, A100}

\bibitem[\protect\citeauthoryear{{Tully}, {Courtois}, {Hoffman}  \&
  {Pomar{\`e}de}}{{Tully} et~al.}{2014}]{tully:2014}
{Tully} R.~B.,  {Courtois} H.,  {Hoffman} Y.,   {Pomar{\`e}de} D.,  2014,
  \mn@doi [\nat] {10.1038/nature13674}, \href
  {https://ui.adsabs.harvard.edu/abs/2014Natur.513...71T} {513, 71}

\bibitem[\protect\citeauthoryear{{Vogeley}, {Hoyle}, {Rojas}  \&
  {Goldberg}}{{Vogeley} et~al.}{2004}]{2004ogci.conf....5V}
{Vogeley} M.~S.,  {Hoyle} F.,  {Rojas} R.~R.,   {Goldberg} D.~M.,  2004, in
  {Diaferio} A.,  ed., IAU Colloq. 195: Outskirts of Galaxy Clusters: Intense
  Life in the Suburbs. pp 5--11 (\mn@eprint {arXiv} {astro-ph/0408583}),
  \mn@doi{10.1017/S1743921304000043}

\bibitem[\protect\citeauthoryear{{Wiegand} \& {Eisenstein}}{{Wiegand} \&
  {Eisenstein}}{2017}]{Wiegand_2017}
{Wiegand} A.,  {Eisenstein} D.~J.,  2017, \mn@doi [\mnras]
  {10.1093/mnras/stx292}, \href
  {https://ui.adsabs.harvard.edu/abs/2017MNRAS.467.3361W} {467, 3361}

\bibitem[\protect\citeauthoryear{Wiegand, Buchert  \& Ostermann}{Wiegand
  et~al.}{2014}]{Wiegand:2013xfa}
Wiegand A.,  Buchert T.,   Ostermann M.,  2014, \mn@doi [Mon. Not. Roy. Astron.
  Soc.] {10.1093/mnras/stu1118}, 443, 241

\bibitem[\protect\citeauthoryear{{Yess} \& {Shandarin}}{{Yess} \&
  {Shandarin}}{1996}]{Yess1996}
{Yess} C.,  {Shandarin} S.~F.,  1996, \mn@doi [\apj] {10.1086/177397}, \href
  {https://ui.adsabs.harvard.edu/abs/1996ApJ...465....2Y} {465, 2}

\bibitem[\protect\citeauthoryear{{Yoshiura}, {Shimabukuro}, {Takahashi}  \&
  {Matsubara}}{{Yoshiura} et~al.}{2017}]{yoshiura:2017}
{Yoshiura} S.,  {Shimabukuro} H.,  {Takahashi} K.,   {Matsubara} T.,  2017,
  \mn@doi [\mnras] {10.1093/mnras/stw2701}, \href
  {http://adsabs.harvard.edu/abs/2017MNRAS.465..394Y} {465, 394}

\bibitem[\protect\citeauthoryear{{Zeldovich}, {Einasto}  \&
  {Shandarin}}{{Zeldovich} et~al.}{1982}]{1982Natur.300..407Z}
{Zeldovich} I.~B.,  {Einasto} J.,   {Shandarin} S.~F.,  1982, \mn@doi [\nat]
  {10.1038/300407a0}, \href
  {https://ui.adsabs.harvard.edu/abs/1982Natur.300..407Z} {300, 407}

\bibitem[\protect\citeauthoryear{{de Lapparent}, {Geller}  \& {Huchra}}{{de
  Lapparent} et~al.}{1986}]{1986ApJ...302L...1D}
{de Lapparent} V.,  {Geller} M.~J.,   {Huchra} J.~P.,  1986, \mn@doi [\apjl]
  {10.1086/184625}, \href
  {https://ui.adsabs.harvard.edu/abs/1986ApJ...302L...1D} {302, L1}

\makeatother
\end{thebibliography}

\appendix

\section{$B_3$ spline smoothing kernel}
\label{app:kern}

As superclusters are searched for as regions with the luminosity density over a certain threshold in a continuous volume of space, we have to convert the spatial positions of galaxies into a density field. 
A standard way is by a kernel sum \citep[][sect. 8.3.2]{davison1997}:
\begin{equation}
    \rho(\mathbf{r}) = \frac{1}{a^3}\sum_{i=1}^N K\left( \frac{\mathbf{r} - \mathbf{r}_i}{a}\right),
    \label{eq:dens}
\end{equation}
where the sum is over all $N$ data points, $\mathbf{r}_i$ are the coordinates, $K(\cdot)$ is the kernel,
and $a$ the smoothing scale. 

\begin{figure}
    \centering
    \includegraphics[]{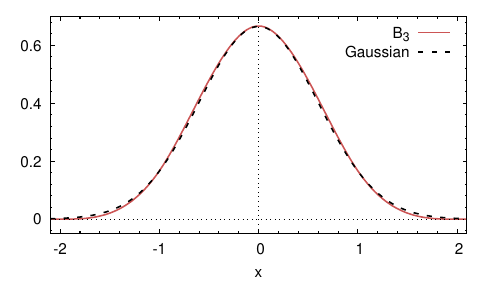}
    \caption{Shape of the $B_3(x)$spline smoothing function. Solid line -- the $B_3(x)$ kernel; dashed line -- a
    Gaussian with $\sigma=0.6$.}
    \label{fig:tuum}
\end{figure}

The kernels $K(\cdot)$ are required to be distributions, positive everywhere and integrating to unity; in our case,
\begin{equation}
    \int K(\mathbf{y})d^3y=1.
    \label{eq:kern}
\end{equation}
Good kernels for calculating densities on a spatial grid are the box splines $B_J$. 
They are local and they are interpolating on a grid:
\begin{equation}
    \sum_i B_J \left(x-i \right) = 1,
    \label{eq:sum}
\end{equation}
for any $x$ and a small number of indices that give non-zero values for $B_J(x)$. 
To create our density fields we use the popular $B_3$ spline function:
\begin{equation}
    B_3(x) = \frac{|x-2|^3 - 4|x-1|^3 + 6|x|^3 - 4|x+1|^3 + |x+2|^3}{12}.
\end{equation}
This function (shown on Figure~\ref{fig:tuum} with comparison to a Gaussian) differs from zero only in the interval $x\in(-2,2)$, meaning that the sum in (\ref{eq:sum}) only includes values of $B_3(x)$ at four consecutive arguments $x\in(-2,2)$ that differ by 1.
The three-dimensional kernel $K_B^{(3)}$ is given by a direct product of three one-dimensional kernels:
\begin{equation}
    K_B^{(3)}(\mathbf{r}) \equiv K_B^{(1)}(x)\, K_B^{(1)}(y)\, K_B^{(1)}(z),
    \label{eq:3dkern}
\end{equation}
where $\mathbf{r} \equiv \{x,y,z\}$. 

\citet{Tempel:2014gr} showed that although $K_B^{(3)}$ is a direct product, it is very close isotropic. They also studied the properties of the density fields calculated using both $B_3$ spline and Gaussian kernels, and found that the density values in high-density regions are practically the same. In low-density environments, however, the differences can be very large with the Gaussian smoothing strongly overestimating density, which can can lead to inaccuracies when studying morphology with Minkowski functionals \citep{saar2009}.

\section{Superclusters found with different fixed thresholds}\label{app:Flat_diff_ths}
\begin{figure}
\centering
\includegraphics[width=\linewidth]{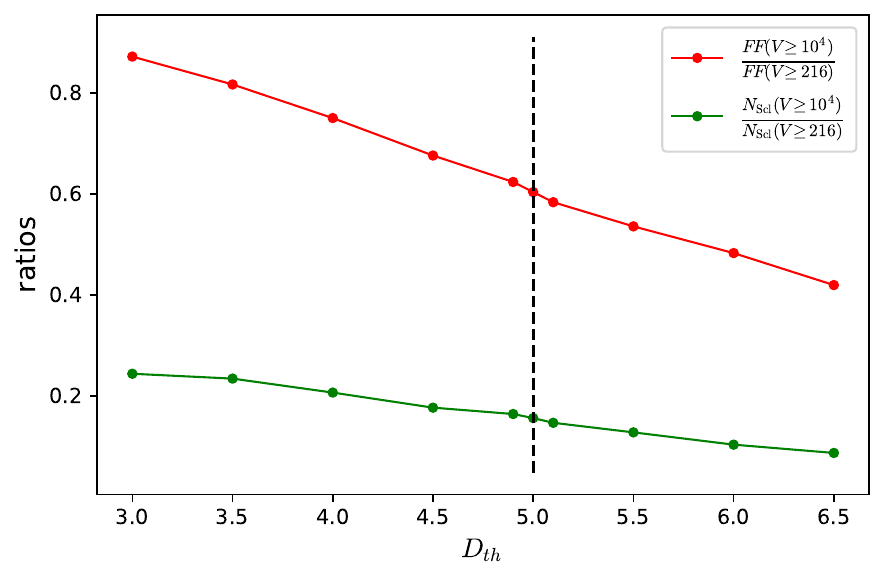}
\caption{The contribution to the filling factor from the large \scls is compared with their number fraction for different fixed thresholds around our primary choice, $\dth=5.0$ shown by the dashed vertical line.
Although, a small fraction of \scls are large but they account for most of the filling factor; e.g. at $\dth=5.0$ only $\sim 18 \%$ of the \scls have volume larger than  $10^4 \mpc^3$ but they enclose $\sim 65\%$ of total supercluster-volume.
}
\label{fig:Dths}
\end{figure}
Supercluster statistics at different fixed thresholds have been presented in Table \ref{tab:flat} in the main text. Note that we define the `fixed threshold' \scls subject to a critical threshold $\dth=5.0$.
 However, we study the properties of \scls defined with slightly different thresholds in this section for comparison. Furthermore, we compare the filling factor (\ff) and the supercluster number (\nc) of the large \scls with that of all the \scls in  Figure~\ref{fig:Dths}. The red and green curves respectively show the ratio of filling factor and \nc between the large and all the \scls as functions of $\dth$. As we raise the threshold, both these ratios decrease as large \scls become rarer. 
Although, a small fraction of \scls are large but they account for most of the supercluster-volume (and hence \ff) across the threshold range in the figure.
For example, at $\dth=5.0$ only $\sim 18 \%$ of the \scls have volume larger than  $10^4 \mpc^3$ but they enclose $\sim 65\%$ of total supercluster-volume.

\subsection{Shapefinders at different fixed thresholds}\label{app:flat_SF_dths}
\begin{figure*}
\centering
\subfigure[$\dth=4.5$]{
\includegraphics[width=0.485\textwidth]{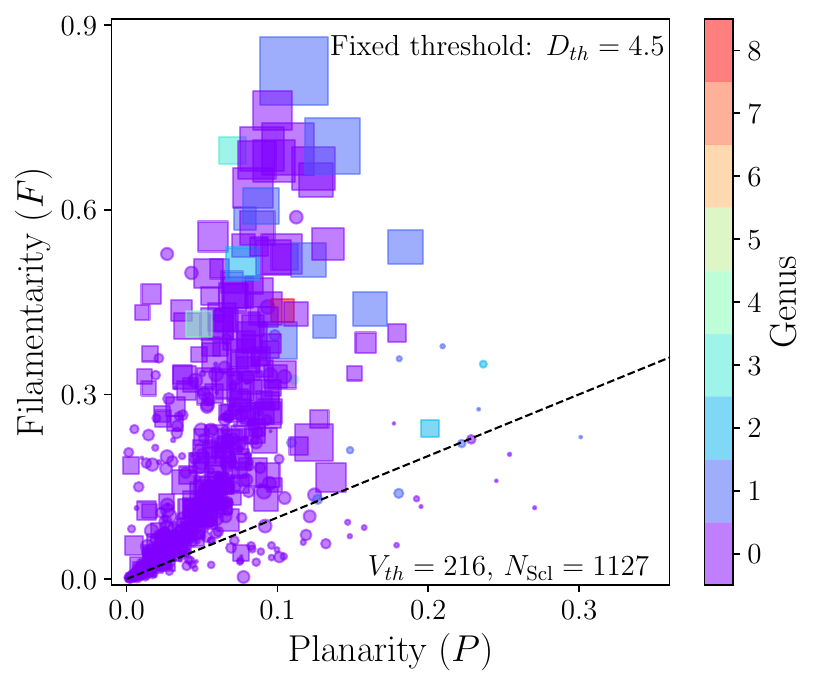}}
\subfigure[$\dth=5.5$]{
\includegraphics[width=0.485\textwidth]{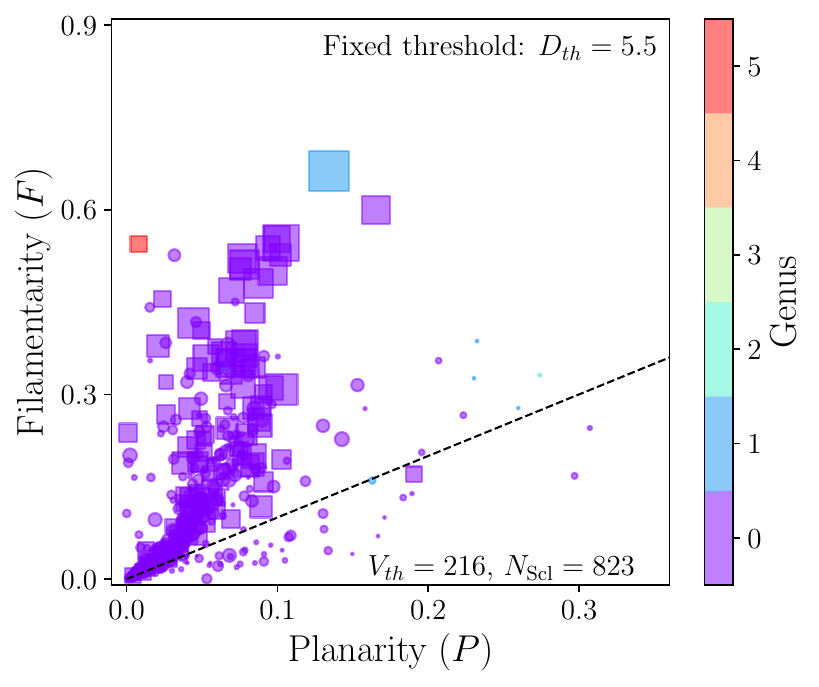}}
\caption{Fixed threshold case: Shapefinders plots for other two values of $\dth$ which can be compared with that of our primary choice, $\dth=5.0$, shown in Figure~\ref{fig:Flat_SF_Dth5}. The smaller ($V < 10^{4}$ Mpc$^3$) and larger ($V \geq 10^{4}$ Mpc$^3$) \scls are marked by the filled circles and squares respectively where the size of the markers is proportional to the volume of the superclusters. The genus value is given by the colour-bar. We again see a few outlier superclusters which tend to have non-zero genus values more often than the others. The dashed lines represent $F=P$ straight lines above which most \scls lie.}
\label{fig:Flat_SF_Dths}
\end{figure*}

Figure~\ref{fig:Flat_SF_Dth5} shows the planarity and filamentarity of individual 
\scls defined with the fixed critical threshold $\dth=5.0$. 
Here, we show similar plots for two slightly different thresholds, 
$\dth=4.5,~5.5$ in Fig.~\ref{fig:Flat_SF_Dths}, for comparison.
At lower threshold density ($\dth=4.5$ shown in the left panel)
we obtain larger number of superclusters,
in agreement with Fig.~\ref{fig:NC}.
Superclusters at this threshold density 
are larger, as they include galaxies from their surrounding regions
with densities $\dth=4.5 - 5.0$. Superclusters also 
  tend to be more filamentary and 
multiply connected than superclusters obtained with the higher threshold
density level ($\dth=5.5$ in the right panel). 
At high threshold density, $\dth=5.5$, galaxies from outer parts of 
superclusters determined with $\dth=5.0$ are excluded, which decreases their 
filamentarity and connectivity.
This explains why higher 
values of filamentarity and genus are found in the \scls 
in the left panel as compared to the right panel.
The overall pattern of the $P-F$ distributions for 
slightly different fixed thresholds 
(comparing Figures~\ref{fig:Flat_SF_Dth5} and \ref{fig:Flat_SF_Dths}) 
is observed to be statistically similar. 

\section{Comparing geometry and shapes of superclusters at different radial distances}\label{app:distance}

\begin{figure*}
\centering
\includegraphics[width=\textwidth]{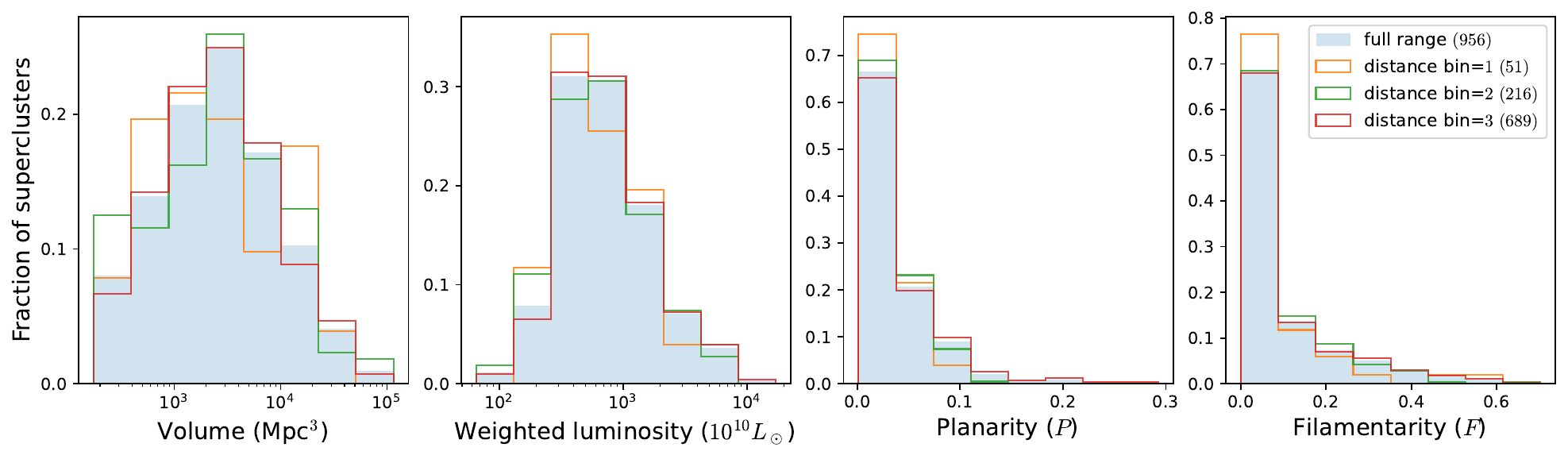}
\caption{The  fixed threshold supercluster sample is divided into three distance bins according to their radial distances from us. The left and right panels in the top row respectively show the volume and luminosity distributions in these the distance bins by the orange, green and red histograms. The bottom-left and bottom-right panels compare the planarity and filamentarity distributions in the superclusters belonging to the three distance bins. The number of supercluster falling to each distance bin has been shown in the parenthesis next to the bin number.  The blue shaded region in all the panels represent the distribution considering all the superclusters.}
\label{fig:dist_comp}
\end{figure*}

In this section we carry out the consistency check that whether the geometrical properties of the fixed threshold \scls remain consistent across the Universe irrespective of their distances from us. For this purpose, we divide the full \scl sample into three bins according to their radial distances. 
The four panels, clockwise from the top-left, of Figure~\ref{fig:dist_comp} compare the distributions of volume, weighted luminosity, filamentarity and planarity of these three sub-samples corresponding to three distance bins. The blue shaded histogram in all the four panels represents the full \scl sample (the number inside the parenthesis next the distance bin number is the number of \scl belonging to that distance bin).

It is evident from the top panels of Figure~\ref{fig:dist_comp} that the \scl sub-samples follow similar volume or weighted luminosity distribution irrespective of their distances from us. However, the nearer distance bin covers smaller SDSS volume. Therefore, it is slightly more difficult to find larger \scls in the nearby universe as evident from the fact that orange histograms in the top panels are slightly skewed towards the smaller volume/luminosity. Since we found strong correlation between the size and the shape of superclusters, the planarity and filamentarity distributions are also quite insensitive to the radial distance as illustrated in the bottom two panels. Therefore, we find that the geometrical properties of the \scls do not vary with the radial distance from us and hence remain consistent across the Universe.

\section{How adaptive threshold superclusters depend on the choice of $\vth$ and $\dthmin$}
\label{app:adaptive}
The methodology of defining \scls with individual thresholds in the adaptive approach is briefly described in section \ref{sec:adaptive_method}. Supercluster sample extracted following this way is most sensitive to the  choices of $\dthmin$ and $\vth$ but depends very little on reasonable choices of $\dthmax$ and $\Delta D$. Some interesting facts regarding the dependencies are given below.
\begin{itemize}
 \item When we decrease $\vth$, keeping $\dthmin$ and $\dthmax$ fixed, we allow more splittings. We note that a new split increases the supercluster number but decreases the filling factor (because after considering the split we also raise the individual threshold for those superclusters). But at the same time, we now allow a few smaller regions near $D \sim \dthmin$ to be considered as superclusters which, on the other hand, raise \ff slightly. 
 This explains why $N_C$ increases but \ff (and \ng) reduces as we decrease $\vth$ while keeping the other important parameters fixed. This characteristic can be observed  in Table \ref{tab:adapt}.
 
 \item On the other hand, if we increase $\dthmax$ only we again permit more splittings, causing an increase in $N_C$ but decay in FF.

 \item Comparing with fixed threshold superclusters: suppose in the fixed threshold case we have maximum $N_C=n_C$ in the interval of threshold $\dthmin \leq D \leq \dthmax$ with a $\vth$. Then in the adaptive case, we will always have $N_C \geq n_C$ with the parameters: $\lbrace \dthmin, \dthmax, \vth \rbrace$. 
 
 \item Raising $\dthmin$ would significantly decrease \ff, as well as $N_C$.  
\end{itemize}
All the above findings are supported by Table \ref{tab:adapt}.

\subsection{Shape distributions for different $\dthmin$ in adaptive threshold approach}\label{app:shape_dist_adap}
\begin{figure*}
\centering
\subfigure[]{
\includegraphics[width=0.485\textwidth]{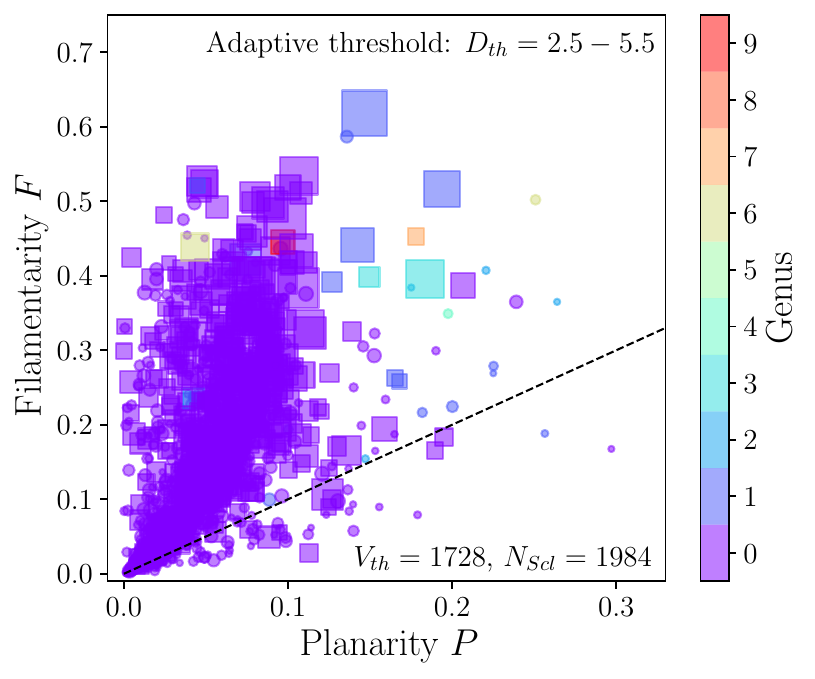}}
\subfigure[]{
\includegraphics[width=0.485\textwidth]{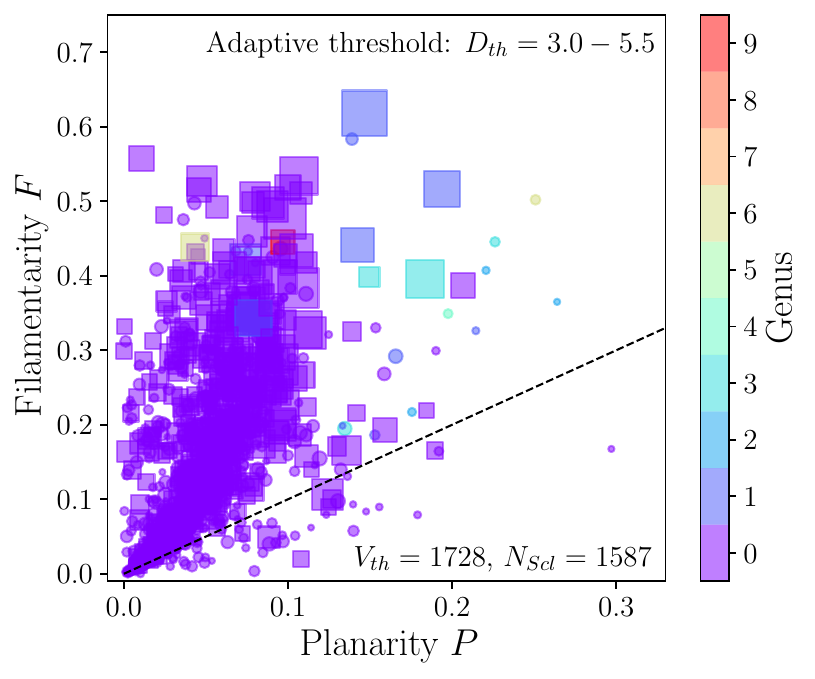}}
\caption{Adaptive threshold case: Shapefinders plots for two other choices of $\dthmin$ which can be compared with that of our primary choice, $\dthmin=3.5$ (shown in Figure~\ref{fig:Adaptive_SF_Dth_main}). The dashed line represents $F=P$ straight line. The general nature of these scatter plots is visibly different (only slightly though) from the corresponding plots for fixed threshold case, compare these plots with those in Figures~\ref{fig:Flat_SF_Dth5}, \ref{fig:Flat_SF_Dths}. }
\label{fig:Adaptive_SF_other_Dths}
\end{figure*}
\begin{figure*}
\centering
\subfigure[]{
\includegraphics[width=0.485\textwidth]{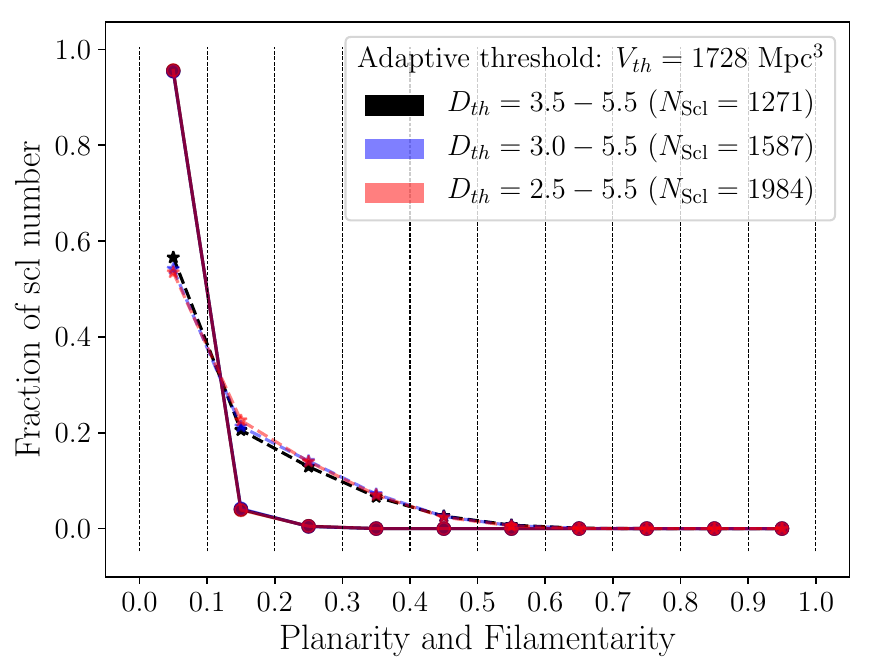}}
\subfigure[]{
\includegraphics[width=0.485\textwidth]{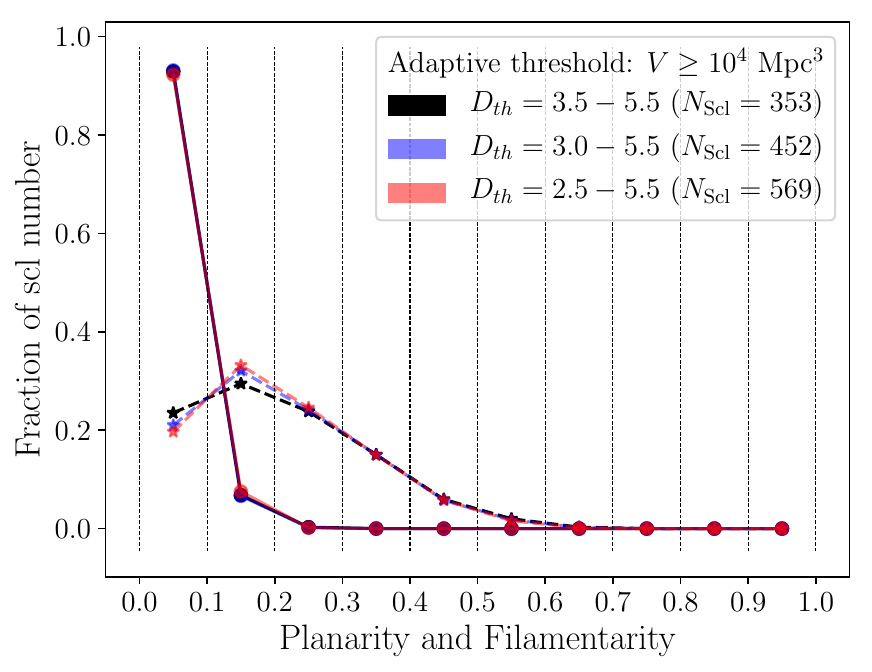}}
\caption{The planarity and filamentarity distributions are shown by the solid and dashed curves for the adaptive threshold \scl samples. Different colours represent various choices of $\dthmin$. The left panel contains the full \scl samples whereas the right panel includes only the large superclusters.}
\label{fig:Adaptive_scl_number_dist}
\end{figure*}

The Shapefinders of individual \scls found in the adaptive threshold method has been shown in Figure~\ref{fig:Adaptive_SF_Dth_main}. We show similar plots for two different choices of $\dthmin$ in Figure~\ref{fig:Adaptive_SF_other_Dths}. All these plots are again comparable to each other, we find more \scls with lower $\dthmin$.

The planarity and filamentarity distributions in \scl samples obtained in adaptive threshold method have been shown in Figure~\ref{fig:Adaptive_scl_number_dist} for three choices of $\dthmin$; the left panel includes all \scls but the right panel focuses on the large \scls with $V \geq 10^4 \mpc^3$. Remarkably, the shape distribution of \scls corresponding to different $\dthmin$ appears to be almost identical in both panels. The majority of the adaptive threshold \scls have planarity smaller than $0.1$. However, close to $22\%$ of the \scls have filamentarity higher than $0.1$. Interestingly, large \scls prefer a non-trivial morphology and the filamentarity distribution exhibits a maximum in the bin $0.1 <F<0.2$ as shown in the right panel. All these observations are consistent with that from Figure~\ref{fig:Adaptive_frac_SF}.

\begin{figure*}
\centering
\subfigure[]{
\includegraphics[width=0.485\textwidth]{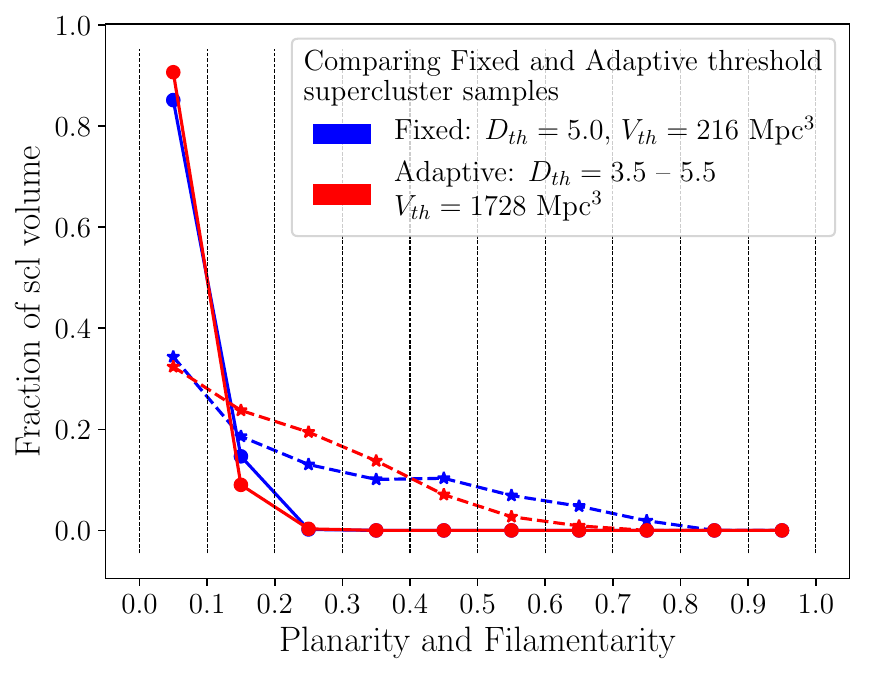}}
\subfigure[]{
\includegraphics[width=0.485\textwidth]{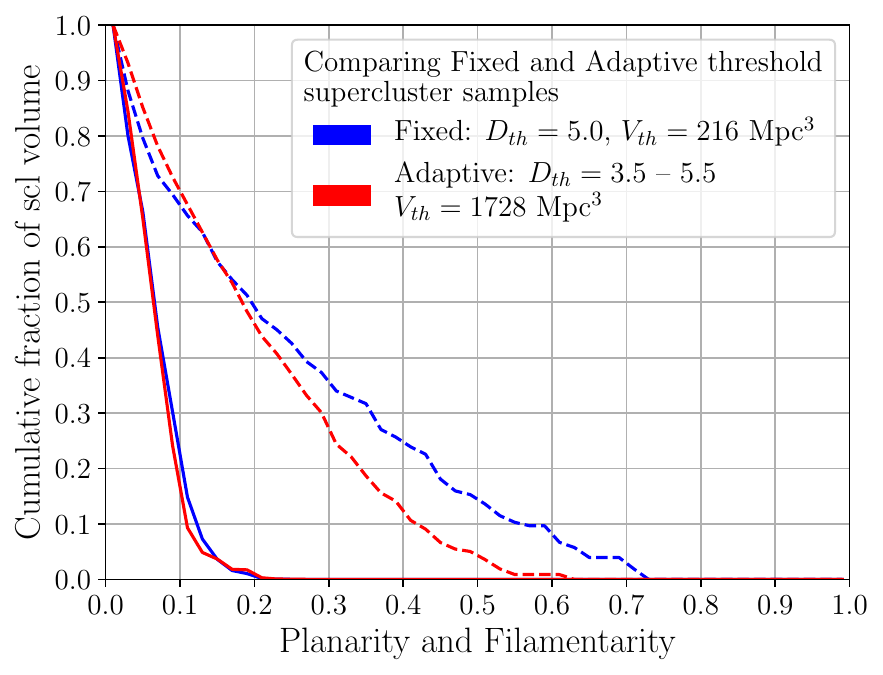}}
\caption{Comparing the shape distribution between the fixed threshold and the adaptive threshold supercluster samples. The solid and dashed curves represent the planarity and filamentarity distribution in both panels. The blue curves correspond to supercluster defined in the fixed threshold approach: $\dth=5.0$. The red curves represent the adaptive threshold superclusters for the choice: $\dthmin=3.5,~\dthmax=5.5, \vth=1728$. We notice that the two definitions of \scls give rise to quite similar shape distributions without any striking difference. However, in the fixed threshold case, slightly larger volume fraction falls inside the higher filamentarity bins whereas in the adaptive threshold case more volume fraction is moderately filamentary.}
\label{fig:Comp_frac_SF}
\end{figure*}

\begin{figure*}
\centering
\subfigure[]{
\includegraphics[width=0.485\textwidth]{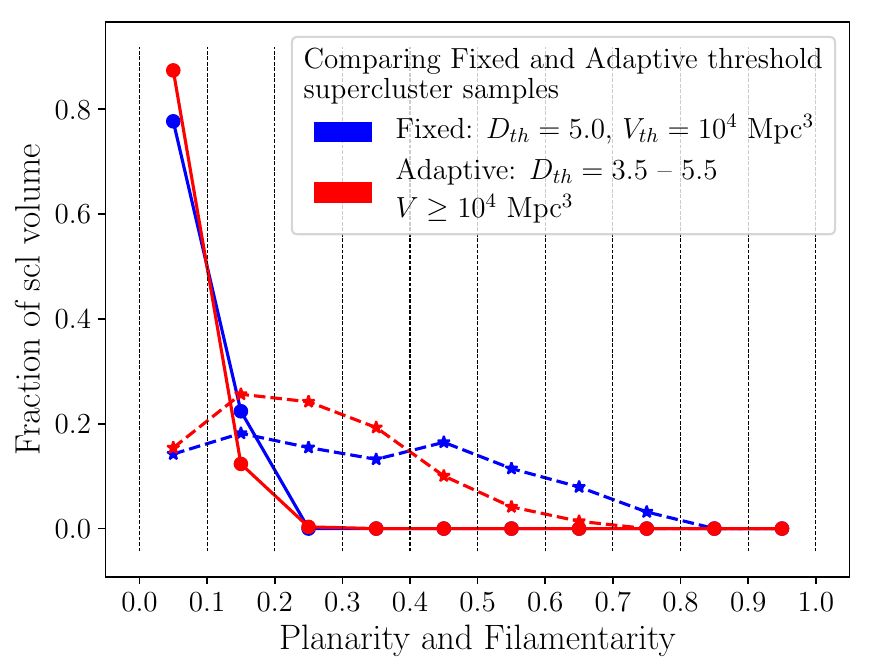}}
\subfigure[]{
\includegraphics[width=0.485\textwidth]{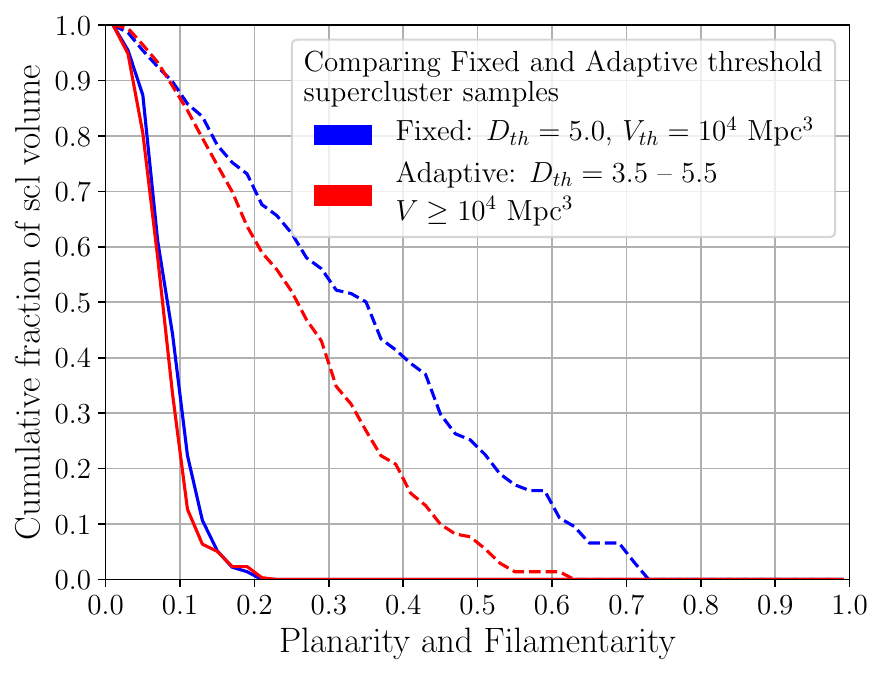}}
\caption{Same as Figure~\ref{fig:Comp_frac_SF} but considering only the large \scls with $V\geq 10^4 ~\mpc^3$.}
\label{fig:Comp_frac_SF_cth10000}
\end{figure*}

\section{Comparing morphology of superclusters obtained in fixed and adaptive threshold methods}
\label{app:comp}

In Figure~\ref{fig:Comp_frac_SF} we compare the shape distributions of \scls defined in fixed and adaptive threshold methods. The plots are already given in Figures~\ref{fig:Flat_frac_SF} and \ref{fig:Adaptive_frac_SF}, we extract the plots corresponding to our baseline thresholds in each case and compare them in Figure~\ref{fig:Comp_frac_SF}. We found no obvious difference in supercluster morphology between the two methods. However, in fixed threshold approach slightly more supercluster-volume is found in the high filamentarity bins whereas in adaptive case more supercluster-volume is in the moderate filamentarity bins. This minor difference becomes more evident when we focus on the large \scls ($V\geq 10^4~\mpc^3$) as shown in Figure~\ref{fig:Comp_frac_SF_cth10000}. The planarity distributions are almost indistinguishable between the two methods.

\section{Superclusters with non-trivial topology ($g>0$)}
\begin{figure*}
\centering
\subfigure[Scl No. 1043: \protect\newline ~~~ $V=1.00 \times 10^{5}~\mpc^3, P=0.15, F=0.62$ \protect\newline ~~~ Origin: (RA, dec, $z$) = ($138.06, 52.18, 0.19$)]{
\includegraphics[width=0.33\textwidth]{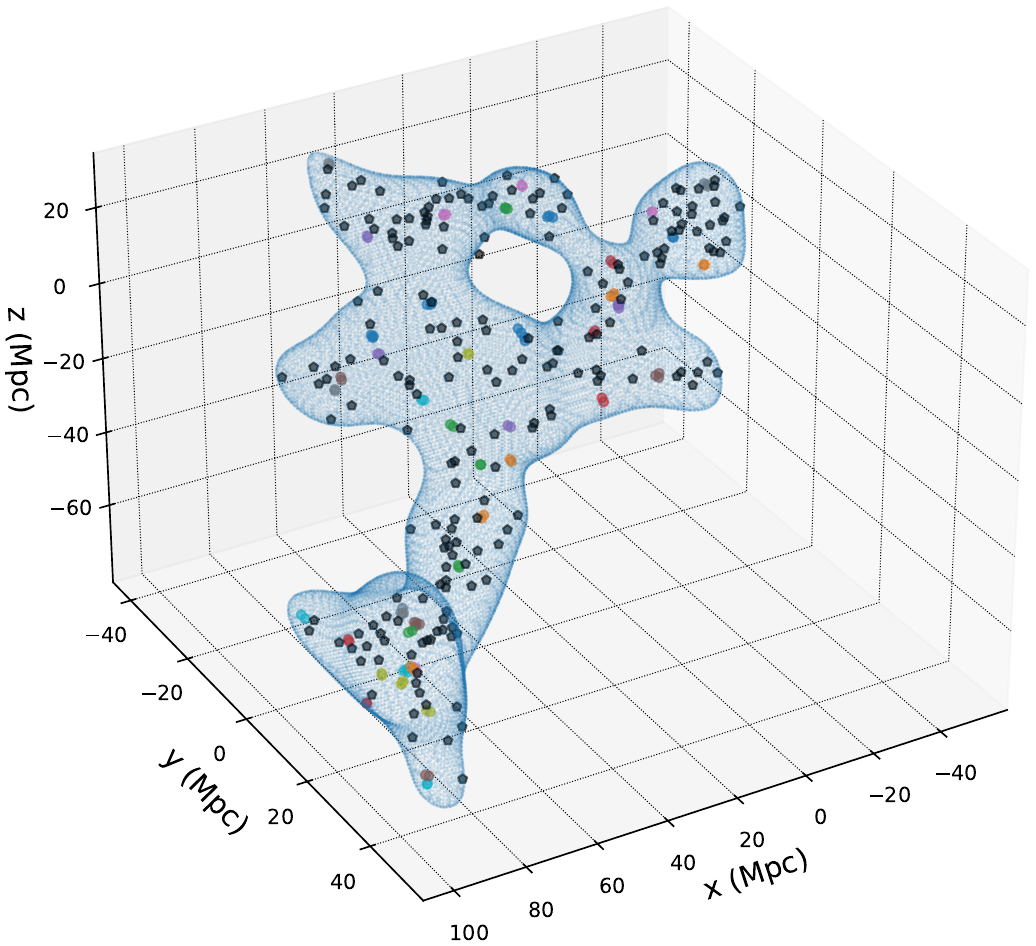}}
\subfigure[Scl No. 864: \protect\newline ~~~ $V=4.74 \times 10^{4}~\mpc^3, P=0.19, F=0.54$ \protect\newline ~~~ (RA, dec, $z$) = ($234.72, 33.64, 0.20$)
]{
\includegraphics[width=0.33\textwidth]{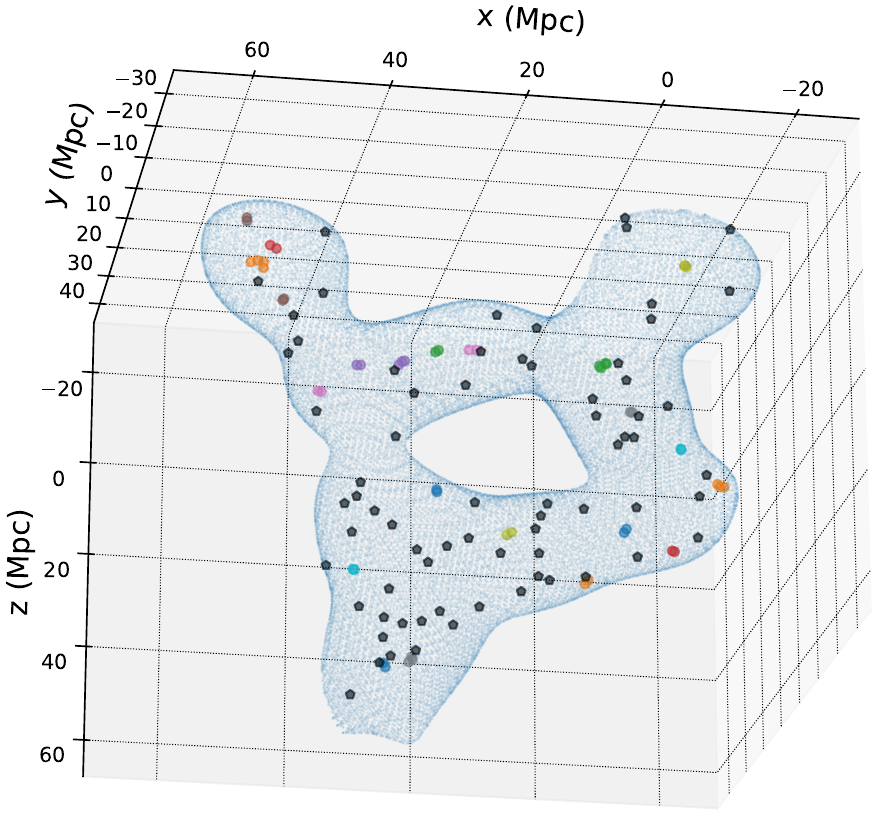}}
\caption{Examples of two superclusters having $G=1$, i.e. one tunnel passes through each of these. The colored filled circles represent the enclosed galaxies belonging to a group whereas the lone galaxies that do not belong to any group are plotted using the black pentagons. Again, the origin is set at the location of the richest group inside each of the superclusters.}
\label{fig:vis_scls_g1}
\end{figure*}

\begin{figure*}
\centering
\includegraphics[width=0.5\linewidth]{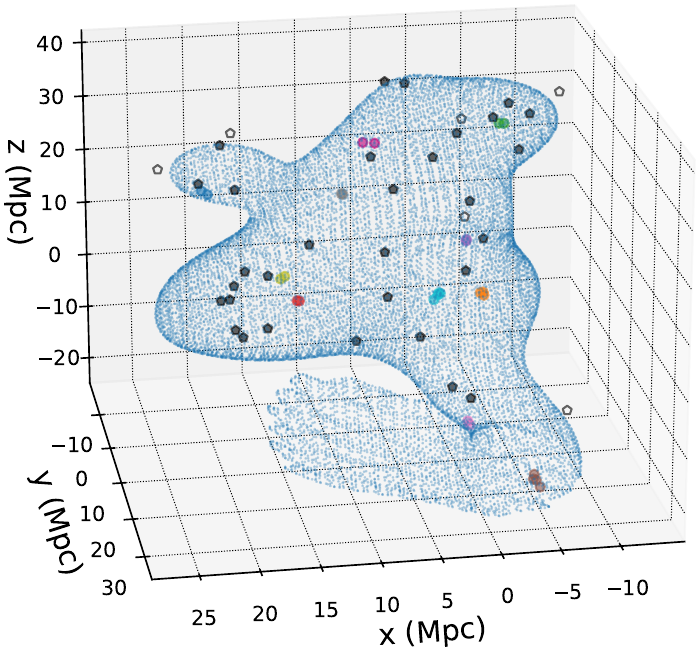}
\caption{Scl No. = 251: $G=8$: $V=1.91 \times 10^{4}~\mpc^3, P=0.05, F=0.53$. (RA, dec, $z$) = ($165.16, 18.07, 0.20$)}.
\label{fig:vis_scls_g8}
\end{figure*}

Although most of \scls (either in fixed or adaptive threshold catalog) have trivial topology (i.e. genus $G=0$), a handful can have non-zero genus $G>0$. Figure \ref{fig:vis_scls_g0} presents 3D visualisation of three \scls with trivial topology. In this appendix, we take a closer look at the \scls with non-trivial topology and see how galaxies are distributed in these. For simplicity, we focus on the \scl catalog found in the fixed threshold $\dth=5.0$ approach where only $1.3\%$ \scls are found to have $G>0$. We find that a few of these \scls have very narrow underdense ($D({\bf r}) < \dth$) tunnels passing through them, but some have very prominent tunnels passing through them. In Figure \ref{fig:vis_scls_g1}, we show two such \scls with $G=1$ in 3D. Similar to Figure \ref{fig:vis_scls_g0}, we adopt here the same Cartesian 3D grid as in our density field but shift the origin of each plot to the location of the richest group whose RA, dec and $z$ are given in the respective panel captions. We can clearly see a prominent underdense tunnel passes through each that gives rise to their non-trivial topology. Both these \scls include significant number of groups and galaxies as given in their panel caption which also mentions the volume, planarity, filamentarity of these superclusters. Note that member galaxies of different groups are shown with different colours and the lone galaxies are marked by black pentagons for each supercluster.

Next we turn our attention to the only \scl that we find to have high genus number (multiply connected, $G=8$ ) in the fixed threshold $\dth=5.0$ catalog. It contains $11$ groups and $58$ galaxies in total as shown in Figure \ref{fig:vis_scls_g8}, the galaxies belonging to different groups are shown by filled circles with different colours and the lone galaxies are marked by filled black pentagons. Note that the eight tunnels passing through this \scl are too narrow to be seen visually in this figure. A close inspection reveals that this particular \scl also appears to be multiply connected with other values of fixed thresholds, and also in the adaptive threshold approach. For example, with fixed threshold $\dth=4.5$ and $5.5$, this \scl have $G=8$ and $5$ respectively (as marked by the red square in both panels of Figure \ref{fig:Flat_SF_Dths}); its size grows with smaller $\dth$ as expected. With our baseline adaptive threshold catalog, this \scl has $G=9$ and it includes the same $11$ groups as in the fixed $\dth=5.0$ catalog. It encloses additional $6$ lone galaxies, which are marked by empty black pentagons, giving rise to $N_{\rm gal}=64$ in total. The key fact that this \scl remains multiply connected in both fixed and adaptive threshold approaches with different parameter values ensures that its multi-connectedness is a robust feature for this given density field. However, since the tunnels are too narrow and this \scl is near the boundary of the SDSS footprint, we are suspicious that the high genus number might arise due to the combination of low number of galaxies and our choice for smoothing scale. Thus we conclude that although we get a handful of \scls with prominent tunnels (e.g. see Figure \ref{fig:vis_scls_g1}), for some, the non-trivial topology can be an artifact.

\end{document}